# CHANDRA/VLA Follow-up of TeV J2032+4131, the Only Unidentified TeV Gamma-ray Source

Yousaf M. Butt[1], Paula Benaglia[2], Jorge A. Combi[2], Michael Corcoran[3], Thomas M. Dame[1], Jeremy Drake[1], Marina Kaufman Bernadó[2], Peter Milne[4], Francesco Miniati[5], Martin Pohl[6], Olaf Reimer[6], Gustavo E. Romero[2], Michael Rupen[7]

[1]Harvard-Smithsonian Center for Astrophysics, 60 Garden St., Cambridge, MA, USA

[2]Instituto Argentino de Radioastronomía, CC 5, 1894, Villa Elisa, Buenos Aires, ARGENTINA

[3] Universities Space Research Association, 7501 Forbes Boulevard, Suite 206, Seabrook, MD 20706 and Laboratory for High Energy Astrophysics, Goddard Space Flight Center, Greenbelt, MD 20771

[4]Los Alamos National Laboratory (T-6),Mail Stop B-277,  Los Alamos, NM 87545, USA

[5]Max-Planck-Institut für Astrophysik, Karl-Schwarzschild-Strasse 1, D-85741 Garching, GERMANY

[6] Institut für Theoretische Physik, Ruhr-Universität Bochum, 44780 Bochum, GERMANY

[7] National Radio Astronomy Observatory, Socorro, NM, USA

**The HEGRA Cherenkov telescope array group recently reported a steady and extended unidentified TeV gamma-ray source lying at the outskirts of Cygnus OB2. This is the most massive stellar association known in the Galaxy, estimated to contain ~2600 OB type members alone. It has been previously argued that the large scale shocks and turbulence induced by the multiple interacting supersonic winds from the many young stars in such associations may play a role in accelerating Galactic cosmic rays. Indeed, Cyg OB2 also coincides with the non-variable MeV-GeV range unidentified EGRET source, 3EG 2033+4118. We report on the near-simultaneous follow-up observations of the extended TeV source region with the CHANDRA X-ray Observatory and the Very Large Array (VLA) radio telescope obtained in order to explore this possibility. Analysis of the CO, HI, and IRAS 100 $\mu$m emissions shows that the TeV source region coincides with an outlying sub-group of powerful OB stars which have evacuated or destroyed much of the ambient atomic, molecular and dust material and which may be related to the very high-energy emissions. An interesting SNR-like structure is also revealed near the TeV source region in the CO, HI and radio emission maps. Applying a numerical simulation which accurately tracks the radio to gamma-ray emission from primary hadrons as well as primary and secondary $e^{\pm}$, we find that the broadband spectrum of the TeV source region favors a predominantly nucleonic – rather than electronic – origin of the high-energy flux, though deeper X-ray and radio observations will help confirm this. A very reasonable, ~0.1%, conversion efficiency of Cyg OB2's extreme stellar wind mechanical luminosity to nucleonic acceleration to ~PeV ($10^{15}$ eV) energies is sufficient to explain the multifrequency emissions.**

I. INTRODUCTION

The astrophysical sites where Galactic cosmic ray (GCR) nuclei gain their extreme energies (up to ~$10^{15}$ eV/nucleon) continue to defy identification. The expanding shock waves of supernova remnants (SNRs) have long been conjectured to be the accelerators of GCRs based mostly on energetic and spectral consistency arguments (eg. Ginzburg & Syrovatskii 1969; Drury et al., 2001). Recent observations from ground-based Cherenkov gamma-ray telescopes have provided direct evidence of TeV range *electrons* in individual SNRs (eg. Muraishi et al, 2000), although the situation for nuclei remains more confused (eg. Reimer & Pohl, 2002; Butt et al 2002; Torres et al., 2003; Erlykin & Wolfendale 2003). Using certain theoretical models it has been possible to interpret the multifrequency emissions from some young SNRs in terms of either nuclear or electron sources, depending on the precise parameters adopted (eg. Gaisser, Protheroe & Stanev, 1998; Ellison, Berezhko & Baring, 2000; Berezhko, Puehlhofer & Völk, 2003).

However, whether or not individual SNRs are sources of GCR nuclei, it is nonetheless important to explore the related (ie. shock driven) acceleration processes thought to operate in conglomerates of SNRs and/or massive stars. Bruhweiler et al. (1980), Kafatos, Bruhweiler and Sofia (1981) among others (eg. McCray & Kafatos 1987; Mac Low & McCray, 1988), have pointed out that since most SNe explosions are core-collapse SNe of massive progenitors (M$\gtrsim$8M$_\odot$), and since such progenitor stars are typically formed in associations, it is plausible that the resultant 'superbubbles' (Heiles, 1979) – characterized by the collective shocks induced by close-by and time-correlated SN explosions – should be even more promising GCR source sites. For recent reviews see, eg., Bykov (2001) and Parizot (2002). From separate considerations of the spallogenic origin of the light elements LiBeB, Ramaty, Lingenfelter, & Kozlovsky (2001) and Alibés, Labay & Canal (2002), also favor the superbubble hypothesis for the origin of GCRs. An important ingredient of such superbubble GCR acceleration models is the additional MHD turbulence induced by the multiple, interacting, supersonic winds blowing from the many young and massive stars present in such associations (eg. Bykov & Fleishman, 1992; Toptygin, 1999; Bykov & Toptygin, 2001).

More than 20 years ago, Cassé and Paul (1980) proposed that the shocked region at the boundary between even a single massive star's stellar wind and the ISM could accelerate nuclei to GCR energies without invoking SNR shocks at all. They pointed out that the integrated mechanical power of a massive star's wind over its lifetime is comparable to the energy liberated in the final SN explosion (~$10^{51}$ ergs). Cesarsky & Montmerle (1983) went further by demonstrating how the turbulent interacting supersonic stellar winds of the many young OB stars in some associations could dominate the GCR acceleration process for the first 4-6 Myrs, even before the first SNe begin to explode. In fact, they suggested that such 'cumulative' OB association stellar winds may be even more efficient than individual SNRs in accelerating GCRs for two reasons: the stellar wind shocks will be turbulent on both sides of the shock interface (thus speeding up the acceleration process); and, since there is continuous energy input, the shock velocity can remain higher for longer than in the impulsively powered SNR shocks.

Of course, it is possible that all 3 shock acceleration processes – among other unrelated mechanisms (eg. Dar & Plaga, 1999) – are responsible for GCR acceleration in varying degrees: individual SNRs (eg. Torres et al., 2003; Erlykin & Wolfendale 2003); correlated SNRs and young stars in superbubbles (eg. Montmerle, 1979; Kafatos, Bruhweiler & Sofia, 1981; Bykov, 2001); and, multiple, interacting, stellar winds in massive OB associations (eg. Cesarsky & Montmerle 1983).

Unfortunately, the direct and firm identification of even a *single* nucleonic GCR acceleration site has continued to elude observers to date. In this context, the recent report by the HEGRA collaboration of an extended and steady TeV source within the boundary of the Cyg OB2 stellar association (Rowell et al., 2002; Aharonian et al. 2002; Horns & Rowell, 2003) provides an ideal opportunity to test the stellar association hypothesis of GCR origin. The low latitude of the source, its ~11 arcmin (Gaussian best-fit) extension, and lack of variability, all point to a Galactic origin[1].

At (4-10) $\times 10^4$ M$_\odot$, Cyg OB2 is the most massive OB association known in the Galaxy; the reader is referred to, eg., Reddish, Lawrence & Pratt (1966); Knödlseder (2000); Comeron et al. (2002); Uyaniker et al. (2001); and, Knödlseder (2002) for useful overviews. Though it houses some of the most massive and luminous stars in the Galaxy – including the only two extreme O3 If* type stars known in the northern hemisphere (stars 7 and 22-A; Knödlseder, 2002) – Cyg OB2 is also a rather compact association: at 1.7 kpc it has a diameter of ~60 pc, or ~2°. This implies a tremendous mechanical power density from the cumulative stellar winds of its ~2600 OB star members: Lozinskaya et al. (2002) estimate that an average of a few $10^{39}$ erg/sec must have been continuously released over the past ~2Myrs in this region.

---

[1] However, the extragalactic alternative cannot be altogether eliminated: an extended *extragalactic* TeV source, the starburst galaxy NGC 253, has been recently reported by the CANGAROO collaboration (Itoh et al., 2002; Itoh et al., 2003) and a possible explanation in terms of cosmic rays illuminating the core regions of massive stars there has been put forth by Romero & Torres (2003) [see also, Anchordoqui, Romero, and Combi, 1999].

Such extreme characteristics make Cyg OB2 a prime candidate for investigating the stellar association hypothesis of the acceleration of GCRs. Already in 1992, White and Chen (1992) predicted that Cyg OB2 ought to be marginally detectable in MeV-GeV gamma-rays by the EGRET instrument based on a model considering the summed $\pi^o \rightarrow \gamma\gamma$ emission from the interactions of energetic nuclei accelerated by just its 4 most luminous members. That the non-variable gamma-ray source, 3EG J2033+4118 (2EG J2033+4112/GRO J2032+40) (Hartman et al. 1999), was found to be centered on Cyg OB2 argues strongly in favor of a physical association (White & Chen, 1992; Chen & White, 1996), although the precise physics of the gamma-ray production may be subject to debate. For instance, it has been argued that the binary system Cyg OB2 #5 may also, by itself, be contributing significant gamma-ray flux by IC upscattering ambient photons from the relativistic electrons known to exist in its colliding wind region (Benaglia et al 2001; Conterras et al., 1997). More broadly speaking, several OB associations are found to be coincident with unidentified EGRET sources, though it is in general difficult to be confident that the associations themselves are the source of the high energy emissions (Romero et al., 1999).

In Figure 1 we show the stellar density plots of all *cataloged* OB member stars together with overlays indicating the positions of 3EG J2033+4118 and TeV J2032+4131 – interestingly, the TeV source coincides with a distinct sub-group of outlying OB stars. Note that many stars in Cyg OB2 remain undetected and uncataloged due to high visual extinction in this direction (eg. Comeron et al., 2002). Six cataloged O, and eight cataloged B stars lie within the reported extent of the TeV source, but again these numbers should be considered strict lower limits. Their parameters and locations are detailed in Table 1.

## II. Observations

The intentions of our follow-up X-ray and radio observations were twofold: firstly, to attempt to identify any likely counterparts of the TeV emission [since, eg., an SNR expanding within hot, low density medium such as an OB association leaves little or no radio/optical signatures (Chu, 1997), X-ray observations can be very enlightening]; and secondly, to measure, or place stringent limits on, the diffuse X-ray and radio emission and thus attempt to constrain whether nuclei or electrons dominate the TeV gamma-ray production.

### a. CHANDRA

We obtained a 5 ksec Director's Discretionary Time (DDT) CHANDRA observation of TeV J2032+4131 ($\alpha_{2000}$: $20^{hr}32^{m}07^{s}\pm9.2^{s}\pm2.2^{s}$, $\delta_{2000}$: $+41°30'30''\pm2.0'\pm0.4'$, radius~5.6'; Aharonian et al., 2002) starting on 11 August 2002 19:51 GMT (OBSID 4358). The data were obtained with the ACIS instrument in very-faint (VF) mode with chips I0,1,2,3 and S2,5. The ~11' TeV source region was centered on the ~16.9' ×16.9' active region of the 4 ACIS-I chips. This field of view comfortably accommodated the ~ ±2' positional error quoted by HEGRA. The data were processed with version 'ASCDS 6.8.0' of the CHANDRA telemetry processing pipelines and were analyzed with CIAO 2.0. A raw (binned-by-8-pixels) image of the ACIS-I chips showing the HEGRA source region is illustrated in Figure 2.

A search for point sources using the *wavdetect* tool resulted in 19 sources above $2.5\sigma$ [15 above $3\sigma$; Table 2]; some associated with already catalogued stars in the region [Table 3]. The source positions have also been overlaid on the ACIS detector image in Figure 3. None of the point sources detected are particularly prominent in X-rays, and none presented sufficient counts to enable detailed spectral analysis. However, since the TeV source is known to be extended (with ~$3\sigma$ confidence) we were particularly interested in investigating the diffuse X-ray emission[2]. We first looked for diffuse structure by adaptively smoothing using the tool *csmooth*[3] an image from which the events associated with the detected point sources had been removed. The result of this smoothing is illustrated in Figure 3, where the detected point sources have been overlaid in green. The diffuse X-ray emission within the region of the TeV source is very weak and shows no significant enhancement over neighboring regions. The smoothed image is brightest toward the southeast of the 5.6' radius HEGRA TeV source region, in the direction of the core of Cyg OB2. We note that the area just northwest of the brightest diffuse region in the southeast corner also tends to harbour most of the detected point sources. A total of 3837 counts (0.3-10 keV in grades 0,2,3,4,6) were detected in the TeV source region, of which 265 can be attributed to point-like sources.

Pulse-height spectra were extracted and telescope response functions calculated for the TeV source region (with point sources removed) using the *acisspec* script. Resulting spectra were analysed using the *sherpa* fitting engine. In order to properly analyse faint spectra of diffuse emitting regions, it is first necessary to account for the particle background that can give rise to significant events in the ACIS detector. A detailed study of the ACIS background has found that, outside of background flare events, both dark moon observations (from which cosmic X-rays are occulted) and observations made with ACIS in the stowed position – out of the focal plane – are characterised by a spectrum of cosmic ray induced events that appears stable over long periods, and that only exhibits relatively small secular changes in overall intensity due to modulation by global solar activity levels (Markevitch et al., 2003). We adopted the methods developed by Markevitch and co-workers to estimate the background based on high signal-to-noise background observations obtained with ACIS in the stowed position[4]. A background spectrum was obtained for the 5.6′ radius HEGRA TeV source region and this was subtracted from the observed spectrum prior to spectral analysis. In addition to this background correction, we also included the affects of the decrease in the quantum efficiency of the ACIS detector as a result of possible filter contamination build-up using the *ACISABS* model[5].

---

[2] Mukerjee et al. (2003) have recently presented a study of this source under the assumption that the TeV emission is *not* extended. However, a recent analysis of new HEGRA data from 2002 have confirmed the extended nature of TeV J2032+4131 at the >5 $\sigma$ level (Horns & Rowell, 2003). It remains possible, however, that *several* distinct point-like TeV sources could be masquerading as a single extended source, given the HEGRA point spread function. Mukerjee and collaborators have also asserted that the possibly associated source 3EG J2033+4118 is variable under the convention of McLaughlin et al. (1996) whereas this source is known to be *non*-variable under all accepted variability schemes, including that of McLaughlin and collaborators (Tompkins, 1999; Torres et al., 2001; Maura McLaughlin, 2003, personal communication; *V*=0.4). There is no indication of source variability beyond the inherent systematics in the method and data itself: it is more than $3\sigma$ from the average AGN variability.

[3] http://cxc.harvard.edu/ciao/ahelp/csmooth.html

Unfortunately, we found that due to the low statistics obtained, the residual TeV source region X-ray spectrum could be equally well-represented by optically-thin plasma models (the *MEKAL* model) or non-thermal power laws. In the case of the former, no constraints were able to be placed on the metallicity parameter: models with metallicity in the range 0-1.2 times the solar photospheric abundances of Anders & Grevesse (1989) were statistically acceptable, yielding reduced $\chi^2$ values of about 0.9. Similar reduced $\chi^2$ values were obtained for power law models. The results of the parameter estimation process for these models are listed in Table 4. The spectrum and model fit for the optically-thin plasma case are illustrated in Figure 4.

Based on the best-fit spectral models, we obtain a diffuse flux within the source region of $1.3 \times 10^{-12}$ ergs cm$^{-2}$ sec$^{-1}$ for the 0.5-2.5 keV bandpass, and $3.6 \times 10^{-12}$ ergs cm$^{-2}$ sec$^{-1}$ for the 2.5-10 keV bandpass. These values are not sensitive to the type of model adopted; power law and optically-thin plasma best-fit models give the same result to within ~5% within the allowed $1\sigma$ parameter ranges for the different models. *Unfortunately, because both power law and thermal plasma models are equally acceptable, the flux values extracted above may only be taken as upper limits to the non-thermal component alone.* Consequently, in our quantitative modeling (Section IV) of the multiwavelength emissions we have taken the measured (instrumental background subtracted) X-ray flux as an upper limit to the X-ray emission associated with the TeV source. A deeper, ~50 ksec, observation would yield sufficient counts to permit a reliable decomposition of the X-ray emission into thermal and power-law components.

---

[4] http://asc.harvard.edu/ciao/threads/acisbackground
[5] http://asc.harvard.edu/ciao/threads/apply_acisabs

Spectra were also extracted for different regions surrounding the TeV source region, including the brighter region to the southeast. The TeV source region showed no significant excess hardness compared to these other regions and spectra were qualitatively very similar.

b. VLA B-configuration

On the following day, 12 August 2002 we obtained a 8 minute 4.86 GHz VLA[6] exposure in the B-configuration, sampling a 10.24′ ×10.24′ region centered at the TeV source (the half-power sensitivity region of the antenna is about 9′ diameter in this configuration). In the B-configuration, the VLA array is sensitive only to point-like radio sources. We achieved an rms noise of $96\mu$Jy/beam for a beam size (psf) of 1.50″ ×1.42″ (FWHM), oriented 28° E of N. We detected no point-like sources to the limiting flux in the region of interest sampled by the primary beam.

c. VLA D-configuration

Since the VLA B-configuration data we obtained is not sensitive to any possible diffuse radio emission present in the TeV source region, we reanalyzed archival D-configuration data at 1.489 GHz taken in 1984 from which we obtained an upper limit to diffuse emission of <200mJy in the region of the TeV source (Figure 5). Our analysis (Section IV) assumes no time variability of the source since 1984, consistent with the multi-year steadiness reported by HEGRA.

d. ROSAT PSPC

We reanalyzed 19.5 ksec ROSAT PSPC data from April/May of 1993 (Sequence # 900314; Waldron et al., 1998). We extracted a source spectrum from a ~12 arcmin diameter circle centered on 20:32:07, +41:30:30, excluding obvious discrete sources. Unfortunately, the PSPC inner support ring runs through this region, which influences the results of our spectral fit. We used a nearby 12' circular region to estimate of background. The net (background subtracted) rate within the TeV source region was 0.107±0.007 PSPC counts/s. An absorbed power law fit yields an acceptable fit with a reduced $\chi^2$ value of 0.72 for 17 degrees of freedom, with a photon index of 0.26, a normalization of $5\times10^{-4}$, NH=0, with a flux (0.2-2.4 keV) of $2\times10^{-12}$ ergs cm$^2$ sec$^{-1}$. A single temperature absorbed thermal model did not yield an acceptable fit (Reduced $\chi^2$ of 2.49 for 17 degrees of freedom). We were able to generate an acceptable fit to the data using a two component thermal model with two separate absorption components (Reduced $\chi^2$ = 0.79 for 14 degrees of freedom). As in the CHANDRA analysis, a hardness image (0.5-2.0 keV) also did not reveal any significant excess hardness in the region of the TeV source. Since our analysis could not resolve the non-thermal vs. thermal nature of the spectrum, the flux $2\times10^{-12}$ ergs cm$^2$ sec$^{-1}$ may be considered an upper limit to the non-thermal emission in the 0.2-2.4 keV band, in good agreement with the CHANDRA results.

---

[6] The VLA is operated by the National Radio Astronomy Observatory (NRAO), which is a facility of the National Science Foundation (NSF), operated under cooperative agreement by Associated Universities, Inc. (AUI).

e. EGRET

The >100 MeV source, 3EG J2033+4118, whose 95% and 99% confidence location contours overlap the extended TeV source region (Fig 1), is a ~12$\sigma$ detection centered at $l$=80.27°, $b$=+0.73°, with a radial positional uncertainty $\theta_{95\%}$=0.28° (Hartman et al. 1999). An elliptical fit by Mattox, Hartman & Reimer (2001) yields the parameters $a$=18.7', $b$=15.0', $\phi$=67°, where $a$ and $b$ are the length of the semimajor and semiminor axes in arcmin, and $\phi$ is the position angle of the semimajor axis in. 3EG J2033+4118 is classified as being a non-variable source by Tompkins (1999), Torres et al. (2001), McLaughlin et al. (1996; $V$=0.61 for 2EG J2033+4112) and M. McLaughlin ($V$=0.4 for 3EG J2033+4118; personal comm., 2003).

At energies above a GeV, the narrower instrumental point spread function of EGRET and the less dominant diffuse gamma-ray background usually enables better source locations for gamma-ray point sources. This is possible if the source spectrum falls less steeply than the spectrum of the diffuse gamma-ray emission above a GeV, and if sufficient photons for an analysis are still available at the higher energies. Two compilations of gamma-ray sources at E>1 GeV have been obtained which differ in subtle, but important, details: the GeV catalog of Lamb & Macomb (1997) and the GRO catalog of Reimer et al. (1997). Only the sources GeV J2035+4214/GRO J2034+4203 from these two catalogs, respectively, could possibly be counterparts for the TeV source position, though it is highly unlikely based on the large positional offsets.

**GeV J2035+4214** (Lamb & Macomb, 1997): $l$=81.22°, $b$=1.02°, detection significance 6.6$\sigma$, and >1 GeV flux (8.1±1.5) × $10^{-8}$ photons cm$^{-2}$ s$^{-1}$; position uncertainties for elliptical fit at 95% contour: $a$=25.4' $b$=17.3' $\phi$=25°

**GRO J2034+4203** (Reimer, Dingus, Nolan, 1997): $l$=80.97º, $b$=1.04º, detection significance 5.8$\sigma$, and >1 GeV flux (5.7±1.3) × $10^{-8}$ photons $cm^{-2}$ $s^{-1}$; 95% and 68% errors of 21' and 14', respectively.

Thus, the 3EG contour fit (E>100 MeV) is actually narrower (E>100 MeV: $a$=18.7', $b$=15.0' $\phi$=67º ) than the one at E > 1 GeV. This is quite unusual and points toward a unfavorable (ie. very soft) spectral index at energies above 1 GeV. In fact, the spectrum of 3EG J2033+4118 has already been studied for representation beyond the single power law fit (index of 1.96±0.1 given in the 3EG catalog) and is significantly better represented if higher order spectral fits are performed. Bertsch et al. (2000) and Reimer & Bertsch (2001) concluded, that in the case of 3EG J2033+4118 a double power law fit or a power law fit with exponential cutoff are more appropriate. This could partially explain the discrepancy between the EGRET flux and the HEGRA flux in a spectral energy distribution (see Fig 3 in Aharonian et al. 2002) – *if the MeV/GeV emission and the newly discovered TeV source are indeed directly related to the same astronomical object in the Cygnus region*. However, such a scenario is highly problematic in that after the index softens in the GeV range it would then have to re-harden to $\sim -1.9$ at the TeV energies observed by HEGRA. In our opinion, such an interpretation appears to be overly contrived.

Thus, while 3EG J2033+4118 and GeV J2035+4214/GRO J2034+4203 may be due to the same object(s), it is unlikely that the TeV source is *directly* related to any of them. 3EG J2033+4118 is probably connected with some subset of the ~2600 OB stars in the core of Cyg OB2, whereas TeV J2032+4131 could be related to the region coincident with an outlying OB sub-group as shown in Fig 1. The sources may, however, still be considered indirectly related if the particles accelerated to GeV energies by the cumulative wind-shocks from the Cyg OB2 core stars, are reaccelerated to TeV energies by the collective wind shocks and turbulence in the region of the outlying OB sub-group. Verifying such a scenario will require deeper multiwavelength observations.

f. OSSE

During the CGRO mission (1991-1999) 11 separate hard X-ray/soft gamma-ray observations of the Cygnus region with the OSSE detector included TeV J2032+4131. However, the field of view of OSSE was 3.8° ×11.4° and even using the earth-occultation technique one cannot resolve sources separated by less than ~0.5°, which happens to be the angular separation of the TeV source from Cyg X-3. The report of a 4.8 hr periodicity in the detected hard X-ray emission in this region by Matz et al.

(1994) argues strongly for its association with Cyg X-3, and not with the TeV source. We also reanalyzed the possible annihilation radiation from the TeV source region in OSSE data, but none was found; the 3-$\sigma$ upper limits being 1.4 $\times 10^{-4}$ ph cm$^{-2}$ sec$^{-1}$ for the 511 keV line and 5.0 $\times 10^{-4}$ ph cm$^{-2}$ sec$^{-1}$ for the positronium continuum. (Care should be taken in comparing these limits with theoretical multiwavelength fits, since most models do not account for annihilation radiation).

## III. The Atomic, Molecular and Dust Morphologies

The distribution of the local diffuse atomic, molecular and dust material is important to understand since it influences the damping and propagation of shocks produced by the stars in Cyg OB2, and can thus provide insight into the distribution and channeling of high-energy particles. It is also crucial to estimate the density of diffuse material in the region of the extended TeV source in order to be able to model the multiwavelength emissions. It should be stressed that distances inferred from gas velocities are very uncertain in the direction of Cyg OB2. Since our line of sight is nearly tangent to the solar circle, radial velocity increases only gradually with distance to a peak of ~4 km s$^{-1}$ at the subcentral distance of 1.4 kpc, then falls back to 0 km s$^{-1}$ at 2.8 kpc, where our line of sight intersects the solar circle. The shallow velocity gradient causes severe blending of emission from the local spiral arm, thought to be viewed tangentially in this direction. Figure 10 of Molnar et al. (1995) provides a very good overview of the Cyg OB2 line-of-sight.

### a. The CO, HI, and ionized Hydrogen Distribution

The CO $J$=1$\rightarrow$0 rotational transition is the best general purpose tracer of molecular hydrogen gas. Using the Galactic CO survey of Dame, Hartmann & Thaddeus (2001), we find good evidence for a molecular gas cavity centered roughly at ($l,b,$ $v_{lsr}$)~(80.5°,+1.8°, +3 km sec$^{-1}$), ~0.8 degrees northwest of the TeV source. The 3 orthogonal slices through the CO $l$-$b$-$v_{lsr}$ data cube shown in Figure 6 suggest that the cavity is the center of an expanding shell with approximate dimensions marked by the dotted ellipses. The $b$-$v_{lsr}$ (Fig. 6a) and $l$-$v_{lsr}$ (Fig. 6c) maps further suggest that a front section of the shell may have been blown out toward us, the remnants of that section seen at $v_{lsr}$ ~ -30 km s$^{-1}$. There are also hints in the $l$-$b$ map (Fig. 6b) of other larger, partial shells roughly centered on Cyg OB2 (mainly in the denser gas at lower latitudes). Using a CO-to-H$_2$ mass conversion factor of 1.8$\times 10^{20}$ cm$^{-2}$ K$^{-1}$ km$^{-1}$ s (Dame, Hartmann, & Thaddeus 2001) and adopting the distance of Cyg OB2 (1.7 kpc), the total H$_2$ mass in the vicinity of the shell ($l$=79° to 81°, $b$=0.5° to 3°, $v_{lsr}$ = −12 to +6 km sec$^{-1}$) is ~3.3

× 10$^5$ M$_\odot$. This value should be considered an upper limit since some emission from unrelated gas in the Local Arm is probably blended in velocity with that from the shell.

We extracted the atomic hydrogen distribution from the Leiden-Dwingeloo HI survey (Burton & Hartmann, 1997), and found a very interesting morphology with respect to the molecular hydrogen traced by the CO data: it appears that the molecular shell encloses a volume of atomic hydrogen, as shown in Figure 7. Note that in this figure the color is the intensity of 21 cm emission integrated -6 to 10 km sec$^{-1}$, and the contours are CO integrated over the same range. The $l$-$v_{lsr}$ and $b$-$v_{lsr}$ maps (Figure 8a,b) demonstrate that the region of enhanced HI fills the CO shell in $l$-$b$-$v_{lsr}$ space. The enhanced HI may be disassociated H$_2$ from the molecular cloud that is currently being overtaken and destroyed by the expanding shell, powered possibly by an SNR or cumulative stellar cluster wind. Interestingly, Langston et al. (2000) have found a number of HII regions distributed on the periphery of this shell-like structure, indicating perhaps that material swept-up by the expansion has triggered star-formation there. Figure 9 displays the CO distribution overlaid on a large-scale 1.42 GHz radio emission map from the Canadian Galactic Plane Survey.

As a rough estimate of the gas density in the region of the TeV source, we calculate here the mean H$_2$ and HI densities within the shell along the line-of-sight to source. Integrating the CO and 21 cm spectra toward the source over the range -4 and +10 km sec$^{-1}$, the estimated velocities of the front and back sides of the CO shell, yields:

N(H$_2$) = 4.2 ×10$^{20}$ H$_2$ cm$^{-2}$ = 8.4 ×10$^{20}$ nucleons cm$^{-2}$
N(HI) = 32.2 × 10$^{20}$ HI cm$^{-2}$

The shell diameter is estimated to be 52 pc, but the path length through the shell along the line of sight to the TeV source is smaller, about 33 pc. Dividing the column densities by this length results in:

n(H$_2$) = 4.1 H$_2$ cm$^{-3}$ = 8.2 nucleons cm$^{-3}$
n(HI) = 31.6 HI cm$^{-3}$

Implicit in this calculation is the assumption that the CO emission over the velocity range -4 to +10 km s$^{-1}$ arises from a real localized object with velocities primarily due to expansion, not Galactic rotation. Otherwise, the velocity range -4 to 10 km s$^{-1}$ would correspond to ~3.7 kpc along the line of sight. Molecular gas is so strongly clumped into large clouds that this assumption is reasonable; indeed, individual GMCs can have internal velocity widths comparable to the full velocity extent of the expanding shell proposed here. On the other hand, the HI gas is much more extended, and some of the 21 cm emission in the velocity range of the shell must be unrelated gas along the line of sight. The HI enhancement which apparently fills the molecular shell does appear superposed on a very substantial background – see, e.g., the color bar in Fig. 7. We estimate that ~65% of the 21 cm emission is actually unrelated to the shell. This reduces the n(HI) value estimated above to 11 cm$^{-3}$ and the mean $H_2$+HI density within the shell to ~19 nucleons cm$^{-3}$.

To this value of density we must also add the density of ionized hydrogen in the region of the TeV source to arrive at an estimate of the total nucleon density. Unfortunately, a precise value for the ionized hydrogen content of the TeV source region alone is not available, but Huchtmeier & Wendker (1977) estimate that there is ~2300 $M_{\odot}$ of ionized hydrogen within the extent of the entire Cyg OB2 association, or ~10 protons cm$^{-3}$ on average.

The mean density of nucleons near the TeV source may then be approximated as: $n_{tot}$($H_2$+HI+proton) ~ 30 nucleons cm$^{-3}$

b. 60 & 100$\mu$m IRAS emission

An examination of the reduced 60 & 100$\mu$m IRAS data (eg. Fig 4b in Odenwald & Schwartz 1993 and Fig 1 in Le Duigou & Knodlseder 2002) clearly shows a dust void at the location of the TeV source. Odenwald & Schwartz (1993) argue that this void is due to the violent stellar environment of Cyg OB2: either the dust has been evacuated from Cyg OB2 – and the TeV source region especially – or else it has been destroyed.

In summary, the molecular and dust maps show a low density region at the location of the TeV source, most plausibly due to the action of the massive core stars of Cyg OB2, as well as the outlying OB sub-group coincident with the TeV source (Fig 1). The co-added atomic+molecular+ionized density of the region of the TeV source is ~30 nucleons cm$^{-3}$.

IV. Modeling the Multifrequency emission.

Determining whether the TeV photons are dominantly produced by electronic or nuclear interactions is, of course, of fundamental importance in assessing whether Cyg OB2 may be considered a nucleonic GCR accelerator. In order to do this, we considered two main cases: one in which the TeV source is due predominantly to $\pi^0 \rightarrow \gamma\gamma$ emission from interactions of energetic nucleons; and the other in which IC upscattering of CMB photons by relativistic electrons generates the bulk of observed gamma-rays. (Considering the measured density of the TeV source region, the IC process will outshine electronic bremsstrahlung in the TeV gamma-ray domain, so we are justified in considering just the two cases mentioned). A third case showing the effects of a lower density electron IC source was also calculated for the purpose of illustration.

We stress that we do not offer here any specific mechanism of accelerating the particles to such high energies since this has been addressed already by several authors, eg., Cesarsky & Montmerle, 1983; Bykov & Fleishman, 1992; Toptygin, 1999; Bykov & Toptygin, 2001; Bykov 2001. We simply assess whether the multiband emissions of the TeV source region are more consistent with a predominantly hadronic *vs.* a predominantly electronic origin, regardless of how the particles may be accelerated to such energies.

To do so we assume that the putative acceleration mechanism (either shock and/or turbulent acceleration) generates a power-law spectrum of primary particles with a normalization, slope and maximum energy chosen to agree with those determined empirically from the observed TeV spectrum. Following Aharonian et al. (2002) we take the spectral index as $-1.9$ and the maximum particle energy as 1 PeV. The required kinetic energy of the injected particles corresponds to only a fraction of a percent of the estimated kinetic energy available in the collective winds of Cyg OB2 (Lozinskaya et al., 2002).

The accelerated particles are assumed to be produced at a constant rate during the lifetime of the source, $\tau_{source}$, which is taken to be ~2.5 Myrs (Knodlseder et al., 2001). The model results depend on the combination of $\tau_{source}$ and environmental parameters (eg. density, magnetic field) that determine the particle losses. However, the impact of the precise value of $\tau_{source}$ is minor: as long as the particles have reached steady-state ($\tau_{cool} < \tau_{source}$ ; as in the leptonic case below), the required energy in the injected particles is proportional to $\tau_{source}$. When $\tau_{cool} \gtrsim \tau_{source}$, as in the hadronic case described below, a reduction in $\tau_{source}$ would require a corresponding increase in the injection power to reproduce the observed radiation. Unlike the leptonic case, the required total energy of the injected particles is left unchanged.

The evolution of the injected particles is followed by integrating a transfer equation (eg. Ginzburg & Syrovatskii 1964) as detailed in Miniati (2001, 2002). For the hadronic component we include losses due to Coulomb collisions, bremsstrahlung and p-p interactions, appropriate for the chosen maximum momentum. And for the leptonic part (*including the secondary $e^{\pm}$'s*) we consider Coulomb collisions, bremsstrahlung, synchrotron and inverse Compton. The thermal gas, CRs and magnetic fields are taken as homogenous and equal to their average values. The radiation field for inverse Compton is dominated by the energy density in the cosmic microwave background and we neglect local contributions of both thermal and non-thermal (eg. synchrotron) origin. In all cases, we assume a spherical source of radius=5.6ʹ (or r~2.77 pc at ~1.7 kpc) and mass=66 $M_{\odot}$ corresponding to the above derived nucleon density of $n_{tot} \sim 30$ cm$^{-3}$. (Except in case III where we consider $n_{tot} \sim 1$ cm$^{-3}$, for illustrative purposes). In lieu of an empirically determined value of the magnetic field in Cyg OB2, we assume a field strength of 5$\mu$G, a nominal Galactic value. *However, we stress that typical magnetic fields in young star forming regions could be significantly higher (eg. Crutcher & Lai, 2002).*

The source term for the secondary electrons and positrons is derived self-consistently based on the evolved CR proton distribution function using the cross sections' model summarized in Moskalenko and Strong (1998). The calculation thus accurately tracks the radio through gamma-ray emission from secondary electrons resulting from the decays of charged muons and kaons produced in hadronic interactions. In particular, the code accounts for the two main secondary production channels: p+p $\rightarrow$ $\pi^{\pm}$ + X and p+p $\rightarrow$ $K^{\pm}$ +X. Their relative contributions to production of the secondary electrons is a function of energy so that the fraction of muons from *K* decay is ~8% at 100 GeV, ~19% at 1 TeV and asymptotically approaches 27% at higher energies. Thus the kaon channel cannot be neglected at the super-TeV energies considered here. The pions and kaons both decay eventually to electrons and positrons in the normal fashion (we do not show neutrinos for simplicity): $\pi^{\pm}$ $\rightarrow$ $\mu^{\pm}$ $\rightarrow$ $e^{\pm}$ ; $K^{\pm}$ $\rightarrow$ $\mu^{\pm}$ $\rightarrow$ $e^{\pm}$ (63.5%); or $K^{\pm}$ $\rightarrow$ $\pi^{o} + \pi^{\pm}$ $\rightarrow$ $\gamma\gamma$ + $\mu^{\pm}$ $\rightarrow$ $e^{\pm}$ (21.2%).

Importantly, we find that the broadband (especially radio) emission from the secondary electrons cannot be ignored, as has often been implicitly assumed in multiwavelength analyses of hadronic gamma-ray production in SNRs, and other proposed GCR sources. This is because the age of the source (2-4×10$^{6}$ years) is much longer than the typical age of SNRs in their GCR acceleration phase (~10$^{4}$ years), and thus significantly more secondaries can accumulate in the source region (since their cooling time is longer than the few Myrs age of the source).

The spectra resulting from our calculations are presented below in figures 10, 11 and 12 for three different cases with parameters as summarized below:

- **Case I**: (predominantly hadronic generation of TeV gamma-rays) – Figure 10

  B=5$\mu$G; E$_{p\_max}$=1 PeV; E$_{e\_max}$=1 PeV; R$_{e/p}$=0.01;

  efficiency, $\eta$~ E$_{CR}$/E$_{kin}$ ~0.08%; density=30 cm$^{-3}$

- **Case II**: (e$^-$ IC generation of TeV gamma-rays) – Figure 11

  B=5$\mu$G; E$_{p\_max}$=1 PeV; E$_{e\_max}$=1 PeV; no protons;

  efficiency, $\eta$~ E$_{CR}$/E$_{kin}$ ~0.2%; density=30 cm$^{-3}$

- **Case III**: (e$^-$ IC generation of TeV gamma-rays; low density case) – Figure 12

  B=5$\mu$G; E$_{p\_max}$=1 PeV; E$_{e\_max}$=1 PeV; no protons;

  efficiency, $\eta$~ E$_{CR}$/E$_{kin}$ ~0.2%; density=1 cm$^{-3}$

In Figure 10 we report the scenario in which the TeV gamma-rays have a hadronic origin. The plot shows the multiband spectra from radio to gamma-ray energies due to synchrotron, bremsstrahlung and inverse Compton emission from primary electrons and secondary $e^{\pm}$, and neutral pion decay generated from p-p inelastic collisions. In this case the emission from primary electrons is shown for comparison and we assume a ratio of electrons to protons at relativistic energies of 0.01. While the TeV spectrum is well reproduced by the hadronic emission, the synchrotron emission due to secondaries generated in the same hadronic processes is below observational upper limits at both radio (1.4 GHz) and X-ray (keV range) frequencies. The predicted radio flux, in particular, is only a factor 2-3 below the observed upper-limit. With an assumed 5$\mu$G magnetic field the particles responsible for the radio-synchrotron emission at 1.4 GHz have a Lorentz factor of order 10$^4$. Given the scaling of the synchrotron emission with magnetic field as B$^{1+\alpha}$, where $\alpha$=0.5 is the spectral index, the magnetic field strength is allowed another factor two or so higher before the radio upper limit is violated. This is, however, the most stringent case as we discuss below. At these energies the timescale for bremsstrahlung losses is slightly shorter than the age of the source, meaning that this portion of the spectrum has basically reached steady state configuration (thick target situation). A lower density would imply a lower rate of production of secondary particles through p-p collision and, therefore, of radio emission, despite the larger fraction of energy that would be radiated as synchrotron instead of bremsstrahlung radiation. In this case, for example, reduction in the density by an order of magnitude would allow a magnetic field as high as 20-30 $\mu$G. Finally, a higher gas density would enhance losses through bremsstrahlung which would show up as a bump in the GeV gamma-ray range of the spectrum. Since this would occur at the expenses of synchrotron emission, a higher magnetic field strength would again be allowed. This can be inferred by considering an approximate scaling for the radio flux as $B^{1.5}/n_{tot}$.

The particle distribution of $e^{\pm}$ is characterized by two breaks at momenta of about 1 GeV/c and 1 TeV/c marking, respectively, the transitions from losses dominated by Coulomb interaction to bremsstrahlung and from bremsstrahlung to synchrotron/inverse Compton mechanism. These breaks, in particular the low-energy one, are actually extended and therefore the spectral transitions are smooth. In addition, the spectra of secondaries start to cut off at momenta of a few TeV/c due to the cut-off in the parent proton spectrum and the average energy of a secondary in p-p inelastic collisions (e.g. Mannheim & Schlickeiser 1994). Notice the given the finite lifetime of the source a steady-state configuration has not been reached for all particle energies. In addition, a fraction of the energy is dissipated through Coulomb collision by particles with momenta below approximately a few GeV/c. This implies that the total luminosity of the secondaries is less than half of that produced by decay of neutral pions.

In Figure 11 we consider the case where the TeV flux arises from electron inverse Compton emission. Thus, as compared to the previous case, we increased the injected population of electrons by a factor more than 200 (hadronic contributions are not shown here for clarity). Since the background gas density and magnetic fields are unchanged with respect to the previous case, the same description of the spectral features applies here as well. It is obvious from the figure that in this case both radio and X-ray upper limits are violated. Particularly, in order to reconcile the predicted and measured radio flux at 1.4 GHz would require a magnetic field at the level of ~1 $\mu$G, which is below the Galactic average.

Finally, Figure 12 shows the same case as in Figure 11 but now – for illustration only – using a lower gas density of 1 $cm^{-3}$. Now the radio emitting particles are not affected by bremsstrahlung losses (thin target) which is reflected in the sharper breaks in the radiation spectra. Due to the limited lifetime of the source (2.4 Myr) the synchrotron emission is increased by only a factor of ~3 which implies only a slightly more stringent upper limit on the magnetic field strength with respect to the previous case.

Clearly, even with the low adopted magnetic field of $5\mu G$, electrons are disfavored as the dominant source of the TeV gamma-rays since both the radio and X-ray upper-limits are violated by the synchrotron emission (Figure 11&12).

It is often stated that a massive and dense cloud is needed to explain the TeV emission as being hadronic in origin. However, there are two main ingredients that determine the hadronic luminosity of a given source: one is indeed the value of the ambient density, but the other is the source's local CR power. We find that the low intensity of this TeV source is easily accommodated by the combination of the empirically determined density of just ~30 nucleons $cm^{-3}$ at the source site and the ~0.1% CR acceleration efficiency (ie. ~$10^{36}$ erg $s^{-1}$ in CRs locally). There is no need to invoke a very massive and/or dense molecular cloud at the TeV source site in order to explain the multiwavelength emissions in terms of p-p interactions.

V. Summary and conclusions

We have carried out follow-up X-ray and radio observations of the extended and steady unidentified TeV source region recently reported by the HEGRA collaboration in Cyg OB2, the most massive OB association known in the Galaxy. The new data taken together with the reexamination of archival radio, X-ray, CO, HI and IRAS data suggest that collective turbulence and large-scale shocks due to the interacting supersonic winds of the ~2600 core OB stars of Cyg OB2, with those of an outlying subgroup of powerful OB stars in Cyg OB2 are likely responsible for the observed very-high-energy gamma-ray emissions (Fig. 1). Since new analysis of 2002 HEGRA data confirm the extended and steady nature of the TeV source (Horns & Rowell, 2003), a blazar-like hypothesis of the origin of the TeV flux, such as that explored by Mukerjee et al. (2003), is now untenable. It is, however, possible that the extended TeV source is actually composed of multiple, nearby steady point-like TeV sources such as may result from a concentration of 'target' stars immersed in an intense CR bath (eg. Romero & Torres, 2003). Higher spatial resolution TeV observations, such as those made possible by HESS, may help in resolving this issue. The suggestion that the TeV source may possibly be associated with Cyg X-3 (eg. Aharonian et al., 2002) is also difficult to reconcile with the fact that Cyg X-3's jets lie at $\lesssim 14°$ to the line-of-sight: the de-projected distance between Cyg X-3 and the TeV source at Cyg X-3's location (~9kpc distant) appears to be too large to support such an hypothesis.

We have carried out detailed simulations of the multifrequency spectra of the extended TeV source and favor a scenario where the TeV gamma-rays are dominantly of a nucleonic, rather than an electronic, origin. A magnetic field of just 5 $\mu$G at the TeV source site would rule against the possibility of an electronic origin of the TeV flux (Fig. 11). Since much higher fields are known to exist in young stellar associations (eg. Crutcher & Lai, 2002), a predominantly hadronic source is favored (Fig. 10). We find no need to invoke a dense and/or massive molecular cloud at the extended TeV source site to explain the multifrequency emissions in terms of accelerated hadrons.

Deeper radio and X-ray observations would be useful in order to separate the non-thermal *vs*. thermal components of the diffuse emissions so that straightforward comparisons to multiwavelength simulations can be made. A determination of the Cyg OB2 magnetic field in this region would also place strong constraints on TeV source models and is highly desirable. Further high-sensitivity infrared observations, such as those already carried out by Comerón et al. (2002), would be very useful in order to make an accurate census of the OB stars towards the highly extincted region of the extended TeV source. Of course, future observations by GLAST and the next-generation of steroscopic Cherenkov telescopes (HESS, VERITAS, etc.) will be critical in exposing the nature of this mysterious very high-energy gamma-ray source.

REFERENCES

1. Aharonian, F. et al., (HEGRA collaboration), An unidentified TeV source in the vicinity of Cyg OB2, A&A 393, L37-L40 (2002)

2. Alibés, A., Labay, J. & Canal, R., Galactic Cosmic Rays from Superbubbles and the Abundances of Lithium, Beryllium, and Boron, 2002 ApJ, 571, 326

3. Anchordoqui, L. A., Romero, G. E.and Combi, J. A., Heavy nuclei at the end of the cosmic-ray spectrum?, 1999 Phys. Rev. D, 60, 103001.

4. Anders, E., & Gravesse, N., Abundances of the elements - Meteoritic and Solar, 1989, Geochimica et Cosmochimica Acta (ISSN 0016-7037), vol. 53, p. 197-214.

5. Benaglia, P., Romero, G. E., Stevens, I. R., Torres, D. F., Can the gamma-ray source 3EG J2033+4118 be produced by the stellar system Cygnus OB2 No. 5?, 2001A&A, 366, 605-611

6. Bertsch, D.L. et al. 2000, Proc. 5th Compton Symposium, AIP 510, 504

7. Berezhko, E. G., Ksenofontov, L. T., Völk, H. J., Emission of SN 1006 produced by accelerated cosmic rays, 2002 A&A., 395, 943

8. Berezhko, E. G., Puehlhofer, G., Völk, H. J, Gamma-ray emission from Cassiopeia A produced by accelerated cosmic rays, Astronomy and Astrophyics, *in press*, astro-ph/0301205

9. Brand, J., Blitz, L., The Velocity Field of the Outer Galaxy, 1993A&A, 275, 67-90

10. Bruhweiler, F. C., Gull, T. R., Kafatos, M., Sofia, S., Stellar winds, supernovae, and the origin of the H I supershells, ApJ L 238, L27-L30

11. Burton, W. B., Hartmann, D., The Leiden/Dwingeloo survey of emission from galactic HI, 1994 Ap&SS, 217, 189

12. Butt, Yousaf M., Torres, Diego F., Romero, Gustavo E., Dame, Thomas M., Combi, Jorge A., Supernova-Remnant Origin of Cosmic Rays?, 2002 Nature 418, 499

13. Bykov, A. M., Fleishman, G. D., On non-thermal particle generation in superbubbles, 1992 MNRAS, 255, 269

14. Bykov, A. M., Toptygin, I. N., A Model of Particle Acceleration to High Energies by Multiple Supernova Explosions in OB Associations, Astronomy Letters, Volume 27, Issue 10, October 2001, pp.625-633

15. Bykov, A., Particle Acceleration and Nonthermal Phenomena in Superbubbles, 2001SSRv, 99, 317-326

16. Cassé, M. and Paul, J. A., Local gamma rays and cosmic-ray acceleration by supersonic stellar winds, 1980ApJ, 237, 236-246

17. Cesarsky, C. J., Montmerle, T., Gamma rays from active regions in the galaxy - The possible contribution of stellar winds, 1983SSRv, 36, 173-193

18. Chen, W., White, R. L., Bertsch, D., Possible detection of $\pi^0$-decay $\gamma$-ray emission from CYG OB2 by EGRET, 1996A&AS, 120C.423-426

19. Chu, Y.-H., Supernova Remnants in OB Assoications, 1997AJ, .113.1815

20. Churchwell, E., in *The Origin of Stars and Planetary Systems*, The Proceedings of the NATO Advanced Study Institute on The Physics of Early Star Formation and Early Evolution, 1998 p.515-552

21. Comerón, F., Pasquali, A., Rodighiero, G., Stanishev, V., De Filippis, E., López Martí, B., Gálvez Ortiz, M. C., Stankov, A., Gredel, R., On the massive star contents of Cygnus OB2, 2002A&A, 389, 874-888

22. Contreras, M. E., Rodriguez, L. F., Tapia, M., Cardini, D., Emanuele, A., Badiali, M., Persi, P., Hipparcos, VLA, and CCD Observations of Cygnus OB2 No. 5: Solving the Mystery of the Radio "Companion", 1997ApJ, 488L.153-156

23. Crutcher, R. M. & Lai, S-P., Magnetic Fields in Regions of Massive Star Formation, in *Hot Star Workshop III: The Earliest Stages of Massive Star Birth. ASP Conference Proceedings, Vol. 267*. Edited by Paul A. Crowther. ISBN: 1-58381-107-9. San Francisco, Astronomical Society of the Pacific, 2002, p.61

24. Dame, T. M., Hartmann, Dap, Thaddeus, P., The Milky Way in Molecular Clouds: A New Complete CO Survey, 2001ApJ, 547, 792-813

25. Dar, A & Plaga, R., 1999 A&A 349, 259

26. Drury, L. O'C. et al., Test of galactic cosmic-ray source models - Working Group Report, 2001SSRv, 99, 329

27. Eichler, D., Usov, V., Particle acceleration and nonthermal radio emission in binaries of early-type stars, 1993ApJ, 402, 271-279

28. Ellison, D., Berezhko, E. & Baring, M., 2000 ApJ, 540, 292

29. Erlykin, A. D., & Wolfendale, A. W., 2003, J.Phys.G: Nucl.Part.Phys, aceepted, astro-ph/0301653

30. Fich, Michel, Blitz, Leo, Stark, Antony A., The rotation curve of the Milky Way to 2 R(0), 1989ApJ, 342, 272-284

31. Gaisser, T., Protheroe, R. & Stanev, T., 1998, ApJ, 492, 219

32. Ginzburg, V. L., Syrovatskii, S. I., The Origin of Cosmic Rays, Topics in Astrophysics and Space Physics, New York: Gordon and Breach, 1969

33. Gosachinskii, I. V., Lozinskaya, T. A., Pravdikova, V. V., The neutral-hydrogen distribution in the region of the radio source Cygnus X, 1999ARep, 43, 391-399

34. Hartman, R. C., Bertsch, D. L., Bloom, S. D., Chen, A. W., Deines-Jones, P., Esposito, J. A., Fichtel, C. E., Friedlander, D. P., Hunter, S. D., McDonald, L. M., Sreekumar, P., Thompson, D. J., Jones, B. B., Lin, Y. C., Michelson, P. F., Nolan, P. L., Tompkins, W. F., Kanbach, G., Mayer-Hasselwander, H. A., Mücke, A., Pohl, M., Reimer, O., Kniffen, D. A., Schneid, E. J., von Montigny, C., Mukherjee, R., Dingus, B. L., The Third EGRET Catalog of High-Energy Gamma-Ray Sources, 1999ApJS, 123, 79-202

35. Heiles, C., HI shells and supershells, 1979ApJ, 229, 533-537

36. Horns, D. & Rowell, G., (for the HEGRA collaboration), in Proceedings of the Second VERITAS symposium on TeV Astrophysics of Extragalactic Sources, Chicago, Il, April 24-26, 2003.

37. Itoh, C., et al., CANGAROO-II collaboration, Detection of diffuse TeV gamma-ray emission from the nearby starburst galaxy NGC 253, A&A, 396 (2002) L1-L4

38. Itoh, C., et al., Galactic Gamma-Ray Halo of the Nearby Starburst Galaxy NGC 253, 2003, ApJL, 584, 65.

39. Kafatos, M., Bruhweiler, F. C. and Sofia, S., Confinement and acceleration of cosmic rays in Galactic superbubbles, Proceedings of the 17[th] International Cosmic Ray Conference, Paris, France, 1981, p.222-225

40. Knödlseder, J., Cygnus OB2 at all wavelengths, Proceedings of the IAU Symposium No. 212, A Massive Star Odessy from Main Sequence to Supernova, *in press*, 2002

41. Knödlseder, J., Cygnus OB2 - a young globular cluster in the Milky Way, 2000A&A, 360, 539

42. Lamb, R.C. and Macomb, D.J. et al. 1997, ApJ 488, 872

43. Langston, G. et al., The First Galactic Plane Survey at 8.35 and 14.35 GHz, 2000 AJ, 119, 2801

44. Le Duigou, J.-M., Knödlseder, J., Characteristics of new star cluster candidates in the Cygnus area, 2002A&A, 392, 869-884

45. Lozinskaya, T. A., Pravdikova, V. V., Finoguenov, A. V., Searches for the Shell Swept up by the Stellar Wind from Cyg OB2, 2002AstL, 28, 223-236

46. MacLow, M.-M. & McCray, M., Superbubbles in disk Galaxies, 1988 ApJ, 324, 776-785

47. Mannheim, K. & Schlickeiser, R., 1994 A&A, 286, 983

48. Markevitch, M., et al., Chandra Spectra of the Soft X-ray Diffuse Background, 2003 ApJ, 583, 70, astro-ph/0209441

49. Massey, P. & Thompson, A. B., Massive stars in CYG OB2, 1991AJ, 101, 1408

50. Mattox, J.R., Hartman, R.C., and Reimer, O. 2001, ApJS 135, 155

51. Matz, S. M., Grabelsky, D. A., Purcell, W. R., Ulmer, M. P., Johnson, W. N., Kinzer, R. L., Kurfess, J. D., Strickman, M. S., Hard X-Ray Variability of Cygnus X-3, The Evolution of X-ray Binaries, Proceedings of a conference held in College Park, MD, 1993. Edited by Steve Holt and Charles S. Day. New York: American Institute of Physics Press. AIP Conference Proceedings, Vol. 308, 1994, p.263-267

52. McCray, R. and Kafatos, M., Supershells and propagating star formation, 1987 ApJ 317, 190-196

53. McLaughlin, M. A., Mattox, J. R., Cordes, J. M. & Thompson, D. J., 1996 ApJ, 473, 763

54. Miniati, F., COSMOCR: A numerical code for cosmic ray studies in computational cosmology, Computer Physics Communications, 2001, Volume 141, Issue 1, p. 17-38.

55. Miniati, F., 2002 MNRAS, 337, 199.

56. Molnar, L. A., Mutel, R. L., Reid, M. J., Johnston, K. J., Interstellar scattering toward Cygnus X-3: Measurements of anisotropy and of the inner scale, 1995ApJ, 438, 708-718

57. Montmerle, T., On gamma-ray sources, supernova remnants, OB associations, and the origin of cosmic rays, 1979, ApJ, 231, 95

58. Moskalenko, I. V., Strong, A. W., Production and Propagation of Cosmic-Ray Positrons and Electrons, 1998ApJ, 493, 694-707

59. Mukerjee, R., Halpern, J. P., Gotthelf, E. V., Eracleous, M., & Mirabal, N., Search for a Point-Source Counterpart of the Unidentified Gamma-Ray Source TeV J2032+4130 in Cygnus, 2003, to appear in ApJ, astro-ph/0302130

60. Muraishi, H. et al. (CANGAROO collaboration), Evidence for TeV gamma-ray emission from the shell type SNR RX J1713.7-3946, 2000A&A, 354L

61. Odenwald, Sten F., Schwartz, Phil R., An IRAS survey of star-forming regions toward Cygnus, 1993ApJ, 405, 706-719

62. Parizot, E., Energetic Particles and Gamma-Rays from Superbubbles, Proceedings of the 2002 Moriond Gamma-Ray Conference, Les Arcs France, *in press*, 2002

63. Prinja, R.K., Barlow, M.J., & Howarth, I 1990, ApJ, 361, 607

64. Ramaty, R., Lingenfelter, R. E., Kozlovsky, B., Spallogenic Light Elements and Cosmic-ray Origin, Space Science Reviews, v. 99, Issue 1/4, p. 51-60 (2001).

65. Reddish, V. C., Lawrence, L. C., Pratt, N. M., The Cygnus II Association, Publ. Royal Observatory Edinburgh, 5, 111-180 (1966)

66. Reimer, O., Dingus, B.L and Nolan, P.L., 1997, Proc. 25th ICRC, Vol.3, 97

67. Reimer, O. & Bertsch, D.L., 2001, Proc. 27th ICRC, p. 2546, astro-ph/0108347

68. Reimer, O., Pohl, M., No evidence yet for hadronic TeV gamma-ray emission from SNR RX J1713.7-3946, 2002A&A, 390L, 43

69. Romero, G. E., Benaglia, P., Torres, D. F., Unidentified 3EG gamma-ray sources at low galactic latitudes, 1999, 1999A&A, 348, 868

70. Romero, G. E. & Torres, D. F., Signatures of hadronic cosmic rays in starbursts? High-energy photons and neutrinos from NGC253, 2003, ApJL accepted, astro-ph/0302149

71. Rowell, G. (for the HEGRA collaboration), An unidentified TeV source in the vicinity of Cyg OB2, Proceedings of the 2002 Moriond Gamma-Ray Conference, Les Arcs France, *in press*, 2002

72. Tompkins, W. F., 1999 PhD Thesis, Stanford University, astro-ph/0202141

73. Toptygin, I. N., Joint particle acceleration by turbulence and a shock front in a steady-state stellar wind, Astronomy Letters, Volume 25, Issue 12, December 1999, pp.814-818

74. Torres, D. F., Romero, G. E., Combi, J. A., Benaglia, P., Anderach, H., and Punsly, B., 2001 A&A 370, 468

75. Torres, D. F., Romero, G. E., Dame, T. M., Combi, J. A., & Butt, Y. M., Supernova remnants and gamma-ray sources, Phys. Rep. 2003, in press, astro-ph/0209565

76. Uyaniker, B., Fürst, E., Reich, W., Aschenbach, B., Wielebinski, R., The Cygnus superbubble revisited, 2001A&A, 371, 675

77. Vacca, W. D., Garmany, C. D., & Schull, J. M. 1996, ApJ, 460, 914

78. Vink, J. S., de Koter, A., & Lamers, H. J. G. L. M., 2000, A&A, 362, 295

79. Waldron, Wayne L., Corcoran, Michael F., Drake, Stephen A.and Smale, Alan P., X-Ray and Radio Observations of the Cygnus OB2 Association, 1998 ApJS, 118, 217

80. Weaver, H., Williams, D. R. W., The Berkeley Low-Latitude Survey of Neutral Hydrogen Part II. Contour Maps, 1974A&AS, 17, 1

81. White, R. and Chen W., Pi0-decay gamma-ray emission from winds of massive stars, 1992ApJ, 387L, 81

We are indebted to Harvey Tannenbaum and the CHANDRA X-ray Center in granting us the DDT time; and to the VLA for allowing the scheduling of the B-configuration observation on such short notice. Discussions with, and information from, Peter Biermann, Dieter Horns, Jurgen Knoldseder, Henric Krawczynski, Maura McLaughlin, Thierry Montmerle, Etienne Parizot, Jerome Rodriguez, Gavin Rowell, Diego Torres, Bulent Uyaniker, Mike Shara, David Thompson, Heinz Wendker and Dave Zurek are appreciated. YMB acknowledges the support of the *CHANDRA* project, NASA Contract NAS8-39073. Study of this extended and non-variable TeV $\gamma$-ray source is also supported by CHANDRA grant DD2-3020X. The use of the HEASARC archive at GSFC, the NASA ADS, and the Canadian Galactic Plane Survey was invaluable to this study.

**Corresponding author Y.M.B. (e-mail: ybutt@cfa.harvard.edu)**

**Table 1.** OB stars surrounding TeV source for $d \leq 9'^{(*)}$

| Name | R.A.$_{2000}$ | Dec$_{2000}$ | Sp.Class.[1] | $r$(pc) | $\log(\dot{M}_{\mathrm{exp}})$ [2] |
|---|---|---|---|---|---|
| Cyg OB2-560 | 20 31 49.74 | 41 28 26.9 | O9.5 V | 2.0 | -6.564 |
| VI CYG 4 | 20 32 13.82 | 41 27 12.0 | O7 III((f)) | 1.7 | -5.567 |
| Cyg OB2-14 | 20 32 16.5 | 41 25 36 | O... | 2.4 | -6.367 |
| Cyg OB2-31 | 20 32 16.62 | 41 25 36.4 | O9 V | 2.4 | -6.367 |
| Cyg OB2-516 | 20 32 25.59 | 41 24 51.9 | O5.5 V | 3.2 | -5.432 |
| Cyg OB2-15 | 20 32 27.5 | 41 26 15 | O8 V | 2.7 | -6.099 |
| Cyg OB2-30 | 20 32 27.66 | 41 26 22.11 | O8 V | 2.7 | -6.099 |
| A43[3] | 20 32 38.5 | 41 25 13.0 | O... | 3.9 | -6.367 |
| MT-299[4] | 20 32 38.66 | 41 25 13.7 | O7.5 V | 3.9 | -6.003 |
| VI CYG 6 | 20 32 45.44 | 41 25 37.51 | O8 V: | 4.2 | -6.099 |
| NSV 13126 | 20 31 22.02 | 41 31 28.4 | B1 Ib: | 4.4 | -5.9 |
| Cyg OB2-205 | 20 31 55.9 | 41 33 04 | B1.5 V | 1.9 | -6.7 |
| Cyg OB2-210 | 20 31 56.4 | 41 31 48 | B1.5 V | 1.2 | -6.7 |
| Cyg OB2-545 | 20 32 03.3 | 41 25 12 | B0.5 :V | 2.4 | -6.4 |
| MT-213 | 20 32 12.8 | 41 22 26 | B0 Vp | 1.5 | -6.2 |
| MT-215 | 20 32 13.2 | 41 27 32 | B1 V | 1.3 | -6.6 |
| Cyg OB2-500 | 20 32 25.8 | 41 29 39 | B1 V | 1.7 | -6.6 |
| Cyg OB2-21 | 20 32 27.4 | 41 28 52 | B1 III | 1.9 | -6.0 |
| Cyg OB2-502 | 20 32 27.85 | 41 28 52 | B0.5 V | 1.9 | -6.4 |
| Cyg OB2-492 | 20 32 36.8 | 41 23 26 | B1 V | 4.3 | -6.6 |

(*): Stars selected from Simbad database, Chen et al. 1996, Comeron et al. 2002, and Massey and Tompson 1991. 1: From Simbad; 2: computed from Vink et al. 2001, if stellar luminosities, masses and effective temperatures are from Vacca et al. 1996, and terminal velocities are from Prinja et al. 1990; 3: Comerón et al. 2002; 4: Massey & Thompson 1991

Table 2: Point-like X-ray sources detected in the TeV source region. The 'SNR' column gives the signal-to-noise ratio of the detection. The netB column refers to the number of counts after background subtraction. The following column, errornetB gives the statistical error in this figure.

| srcid | x | y | errorx | error y | ra | dec | SNR | netB | errornetB |
|---|---|---|---|---|---|---|---|---|---|
| XS043358B0_004 | 5055.43 | 4009.50 | 0.85 | 0.87 | 20:31:23.55 | 41:29: 48.7 | 3.2 | 19 | 6 |
| XS043358B2_007 | 4825.85 | 3648.26 | 0.92 | 0.59 | 20:31:33.63 | 41:26: 51.2 | 3.2 | 17 | 5 |
| XS043358B2_009 | 4743.82 | 3250.92 | 1.41 | 1.17 | 20:31:37.24 | 41:23: 35.7 | 3.2 | 19 | 6 |
| XS043358B6_004 | 4732.06 | 2001.42 | 2.74 | 2.49 | 20:31:37.83 | 41:13: 21.0 | 10.0 | 146 | 15 |
| XS043358B0_002 | 4592.83 | 4753.28 | 0.32 | 0.35 | 20:31:43.77 | 41:35: 55.0 | 12.8 | 190 | 15 |
| XS043358B2_005 | 4422.18 | 3223.85 | 0.77 | 0.40 | 20:31:51.31 | 41:23: 22.6 | 5.1 | 38 | 7 |
| XS043358B0_001 | 4408.75 | 4191.85 | 0.21 | 0.23 | 20:31:51.86 | 41:31: 18.8 | 3.7 | 22 | 6 |
| XS043358B1_004 | 4304.17 | 4932.17 | 0.59 | 0.85 | 20:31:56.43 | 41:37: 23.1 | 4.7 | 33 | 7 |
| XS043358B2_002 | 4067.34 | 3443.97 | 0.53 | 0.37 | 20:32:06.82 | 41:25: 10.9 | 4.2 | 26 | 6 |
| XS043358B3_006 | 3907.58 | 3690.17 | 0.24 | 0.24 | 20:32:13.81 | 41:27: 12.0 | 4.4 | 29 | 6 |
| XS043358B6_005 | 3884.06 | 1785.92 | 4.19 | 3.18 | 20:32:14.80 | 41:11: 35.2 | 4.4 | 41 | 9 |
| XS043358B3_001 | 3792.98 | 3975.72 | 0.16 | 0.21 | 20:32:18.83 | 41:29: 32.5 | 4.1 | 25 | 6 |
| XS043358B3_012 | 3592.20 | 3587.60 | 0.65 | 0.30 | 20:32:27.60 | 41:26: 21.5 | 3.1 | 17 | 5 |
| XS043358B3_019 | 3184.15 | 3495.80 | 1.25 | 1.73 | 20:32:45.45 | 41:25: 36.0 | 2.5 | 15 | 6 |
| XS043358B3_018 | 3031.55 | 3819.36 | 1.57 | 1.56 | 20:32:52.16 | 41:28: 15.0 | 3.0 | 19 | 6 |
| XS043358B3_017 | 2956.44 | 3458.72 | 1.46 | 1.04 | 20:32:55.41 | 41:25: 17.5 | 3.3 | 25 | 8 |
| XS043358B9_002 | 1648.67 | 2734.17 | 0.28 | 0.30 | 20:33:52.45 | 41:19: 18.6 | 2.7 | 23 | 8 |
| XS043358B9_001 | 1643.83 | 2767.17 | 0.27 | 0.27 | 20:33:52.67 | 41:19: 34.8 | 2.7 | 23 | 9 |
| XS043358B9_003 | 1630.75 | 2768.00 | 0.22 | 0.25 | 20:33:53.24 | 41:19: 35.2 | 2.6 | 22 | 8 |

Table 3: Cataloged stars (*) coincident with, or nearby, the point-like X-ray sources listed in Table 4. The spectral type is given when available. The two columns *r=15″* and *r=30″* give the search radius around each X-ray source. Some of the X-ray sources without counterparts may be young stars which have yet to be optically identified due to high extinction towards the Cygnus direction. X stands for previously detected X-ray source.

| srcid | $r = 15''$ | $r = 30''$ | ra | dec | object |
|---|---|---|---|---|---|
| XS04358B0_004 | – | – | – | – | — |
| XS04358B2_007 | – | – | – | – | — |
| XS04358B2_009 | [MT91] 115 | [MT91] 115 | 20 31 37.38 | 41 23 35.5 | * |
| | Ass Cyg OB 2-581 | Ass Cyg OB 2-581 | 20 31 37.4 | 41 23 35 | * |
| | | Ass Cyg OB 2-580 | 20 31 37.2 | 41 23 35 | * |
| | | [MT91] 114 | 20 31 37.20 | 41 23 55.4 | * |
| XS04358B6_004 | NSV 13129 | NSV 13129 | 20 31 37.50 | 41 13 21.2 | *; O9: |
| | [TSA98] J20... | [TSA98] J20... | 20 31 38.83 | 41 13 24.7 | X |
| | | 1RXS J203... | 20 31 40.10 | 41 13 19.0 | X |
| XS04358B0_002 | Ass Cyg OB 2-195 | Ass Cyg OB 2-195 | 20 31 43.8 | 41 36 07 | * |
| | [MT91] 136 | [MT91] 136 | 20 31 43.72 | 41 36 07.6 | * |
| XS04358B2_005 | Ass Cyg OB 2-551 | Ass Cyg OB 2-551 | 20 31 51.4 | 41 23 23 | * |
| | [MT91] 152 | [MT91] 152 | 20 31 51.42 | 41 23 23.6 | * |
| | 2E 2031.1+4112 | 2E 2031.1+4112 | 20 31 50.9 | 41 23 19 | X |
| XS04358B0_001 | [MT91] 150 | [MT91] 150 | 20 31 50.91 | 41 31 17.5 | * |
| | Ass Cyg OB 2-208 | Ass Cyg OB 2-208 | 20 31 50.9 | 41 31 18 | * |
| | | [MT91] 162 | 20 31 54.48 | 41 31 13.5 | * |
| XS04358B1_004 | | Ass Cyg OB 2-197 | 20 31 53.9 | 41 37 30 | * |
| XS04358B2_002 | | [MT91] 206 | 20 32 09.00 | 41 25 05.6 | * |
| | | Ass Cyg OB 2-546 | 20 32 09.1 | 41 25 05 | * |
| XS04358B3_006 | VI CYG 4 | VI CYG 4 | 20 32 13.82 | 41 27 12.0 | *; O7III((f)) |
| | | [MT91] 213 | 20 32 12.8 | 41 27 26 | *; B0Vp |
| | | [MT91] 215 | 20 32 13.2 | 41 27 32 | *; B1V |
| | | [MT91] 221 | 20 32 14.3 | 41 27 41 | * |
| XS04358B6_005 | – | – | – | – | – |
| XS04358B3_001 | | Ass Cyg OB 2-532 | 20 32 19.5 | 41 30 00 | * |
| | | [MT91] 233 | 20 32 19.35 | 41 30 01.0 | * |
| XS04358B3_012 | Ass Cyg OB 2-30 | Ass Cyg OB 2-30 | 20 32 27.66 | 41 26 22.1 | *; O8V |
| | Ass Cyg OB 2-15 | Ass Cyg OB 2-15 | 20 32 27.5 | 41 26 15 | *; O8V |
| XS04358B3_019 | BD+40 4221 | BD+40 4221 | 20 32 45.4 | 41 25 37 | * |
| | VI CYG 6 | VI CYG 6 | 20 32 45.44 | 41 25 37.5 | *; O8V: |
| | | [MT91] 312 | 20 32 43.61 | 41 25 38.4 | * |
| | | Ass Cyg OB 2-483 | 20 32 43.6 | 41 25 38 | * |
| XS04358B3_018 | [MT91] 351 | [MT91] 351 | 20 32 52.87 | 41 28 19.2 | * |
| | Ass Cyg OB 2-477 | Ass Cyg OB 2-477 | 20 32 52.9 | 41 28 19 | * |
| XS04358B3_017 | [MT91] 360 | [MT91] 360 | 20 32 54.88 | 41 25 15.6 | * |
| XS04358B9_002 | – | – | – | – | – |
| XS04358B9_001 | – | – | – | – | – |
| XS04358B9_003 | – | – | – | – | – |

Table 4: Details of the model parameters used to fit the background subtracted diffuse X-ray spectrum in the TeV source region. Due to poor statistics we cannot constrain the nature of the emission: thermal *vs.* power-law. Both model fits yield approximately the same reduced $\chi^2{\sim}0.9$.

| Model | Parameter | Best-Fit | $1\sigma$ Range | |
|-------|-----------|----------|---------|---|
| *Optically–* | kT (KeV) | 11.3 | -3.3 | +5.7 |
| *thin plasma* | Abundance | < 1.2 (1$\sigma$) | | |
| | normalisation* | $2.84\times10^{-3}$ | $-0.4\times10^{-3}$ | $+0.4\times10^{-3}$ |
| | NH (cm$^{-2}$) | $1.5\times10^{21}$ | $-0.4\times10^{21}$ | $+0.4\times10^{21}$ |
| *Power Law* | Photon Index | 1.53 | -0.11 | +0.12 |
| | normalisation** | $7.6\times10^{-4}$ | $-0.8\times10^{-4}$ | $+1.0\times10^{-4}$ |
| | NH | $1.8\times10^{21}$ | $-0.5\times10^{21}$ | $+0.5\times10^{21}$ |

Units for normalisation:

$$* \frac{10^{-14}}{4\pi D^2}\int n_e n_H dV \quad \text{where } D \text{ is the distance}$$

** *photons keV$^{-1}$ cm$^{-2}$ sec$^{-1}$* at 1 keV

<u>Fluxes</u>

```
Flux (0.5,2.5) keV = 0.0006 photons cm⁻² sec⁻¹
Flux (0.5,2.5) keV = 1.4×10⁻¹² ergs cm⁻² sec⁻¹

Flux (2.5,10) keV = 0.00045 photons cm⁻² sec⁻¹
Flux (2.5,10) keV = 3.6×10⁻¹² ergs cm⁻² sec⁻¹
```

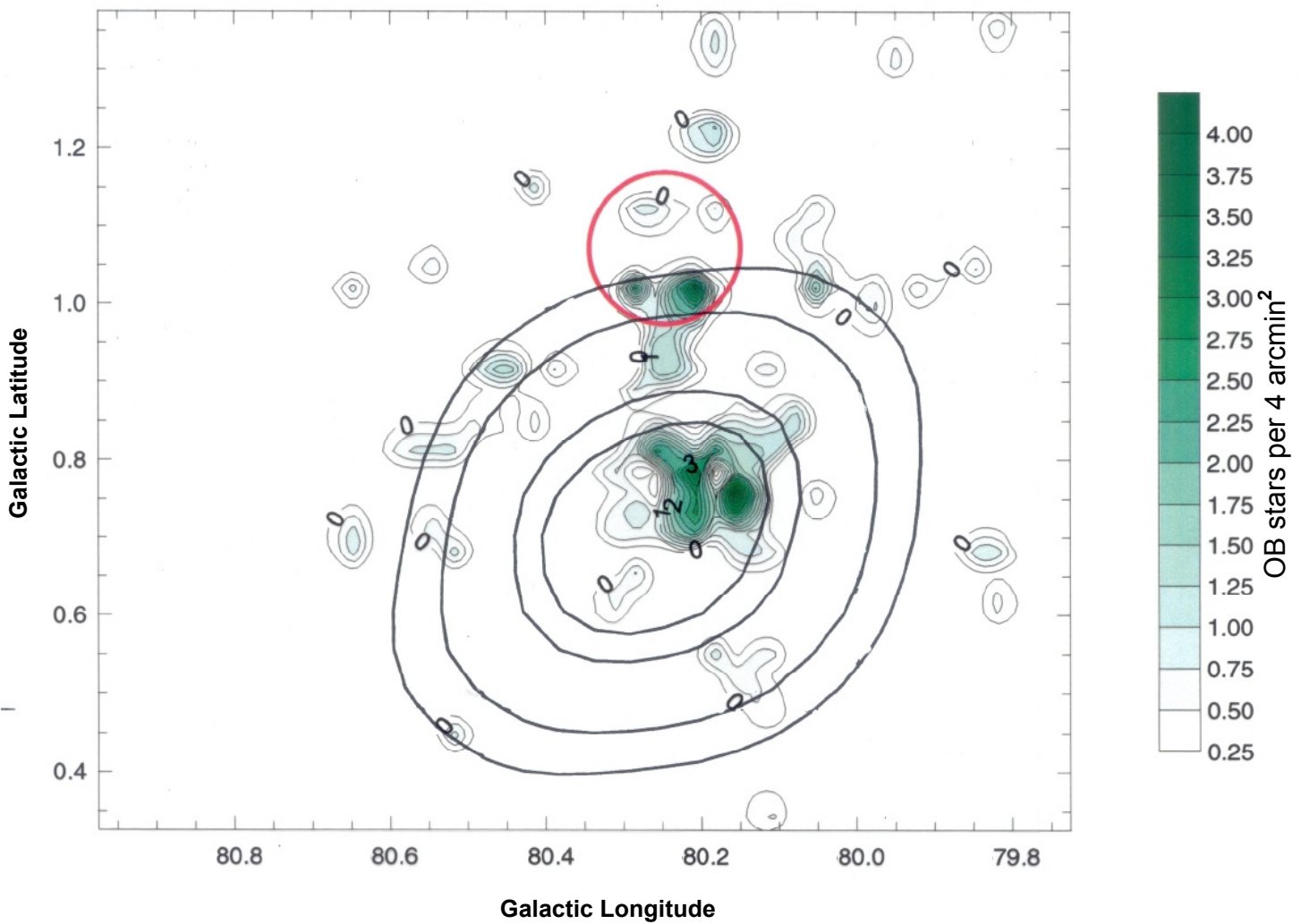

Fig. 1: Distribution of all 110 *cataloged* OB stars in Cyg OB2 shown as a surface density plot (stars per 4 arcmin$^2$). Note that many stars in Cyg OB2 remain uncataloged – the total number of OB stars alone is expected to be ~2600 (Knodlseder 2002). Although the extinction pattern towards Cyg OB2 may control the observed surface density of OB stars, our analysis assumes that the observed distribution of OB stars tracks the actual distribution. The thick contours show the location probability contours (successively, 50%, 68%, 95%, and 99%) of the non-variable MeV-GeV range EGRET $\gamma$-ray source 3EG 2033+4118 (Hartman et al., 1999). The red circle outlines the 5.6' radius extent of the diffuse and steady TeV source, TeV J2032+4131, reported by HEGRA (Rowell et al. 2002; Aharonian et al., 2002)

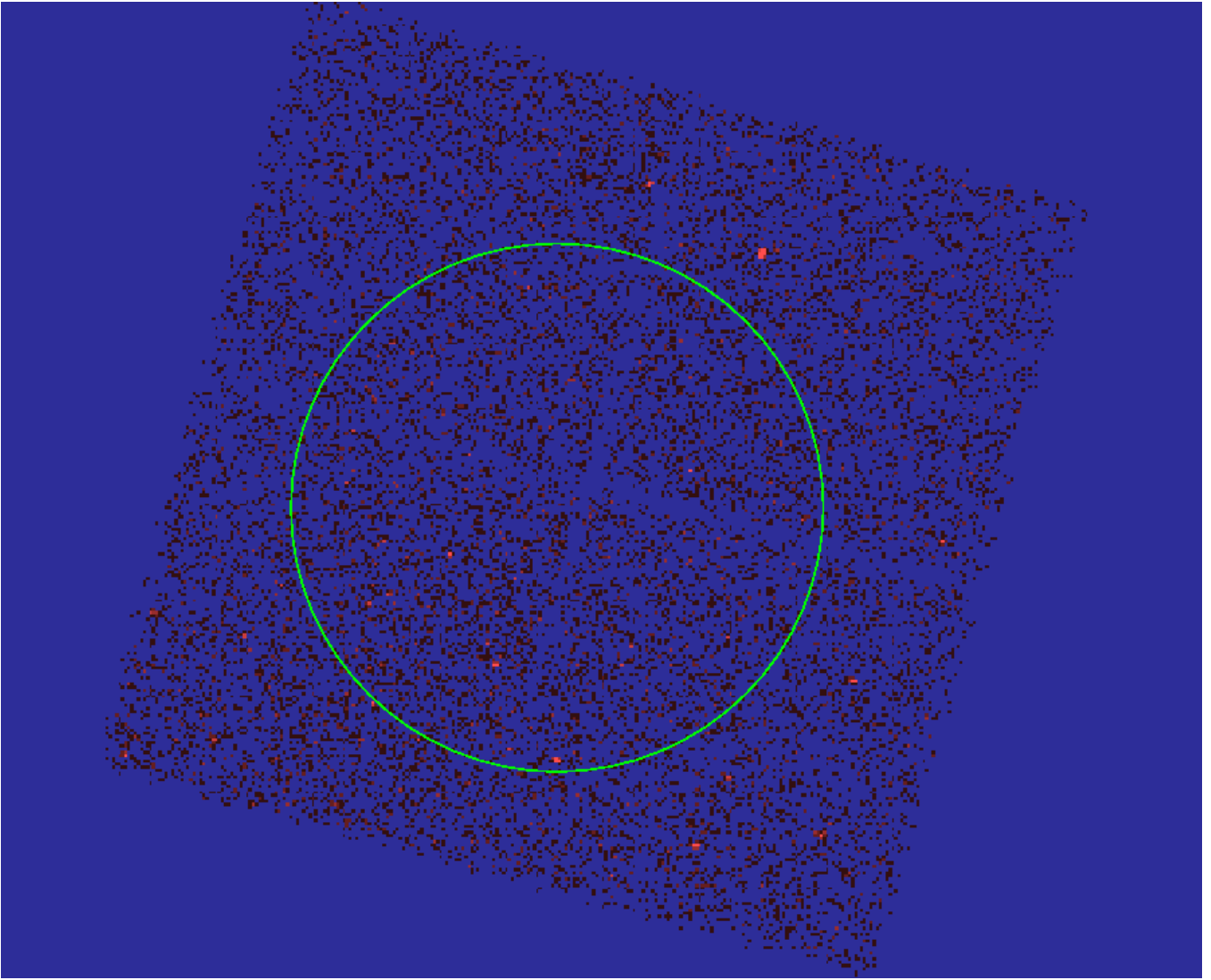

Fig. 2: The raw 5 ksec CHANDRA image of the 4 I-array chips (binned-by-8-pixels). The green circle shows the 5.6′ radius extent of the diffuse TeV source, TeV J2032+4131, reported by HEGRA (Aharonian et al., 2002). The aimpoint is at the center of the circle, $\alpha_{2000}$: $20^{\mathrm{hr}}32^{\mathrm{m}}07^{\mathrm{s}}$, $\delta_{2000}$: $+41°30′30″$. North is up and East is to the left.

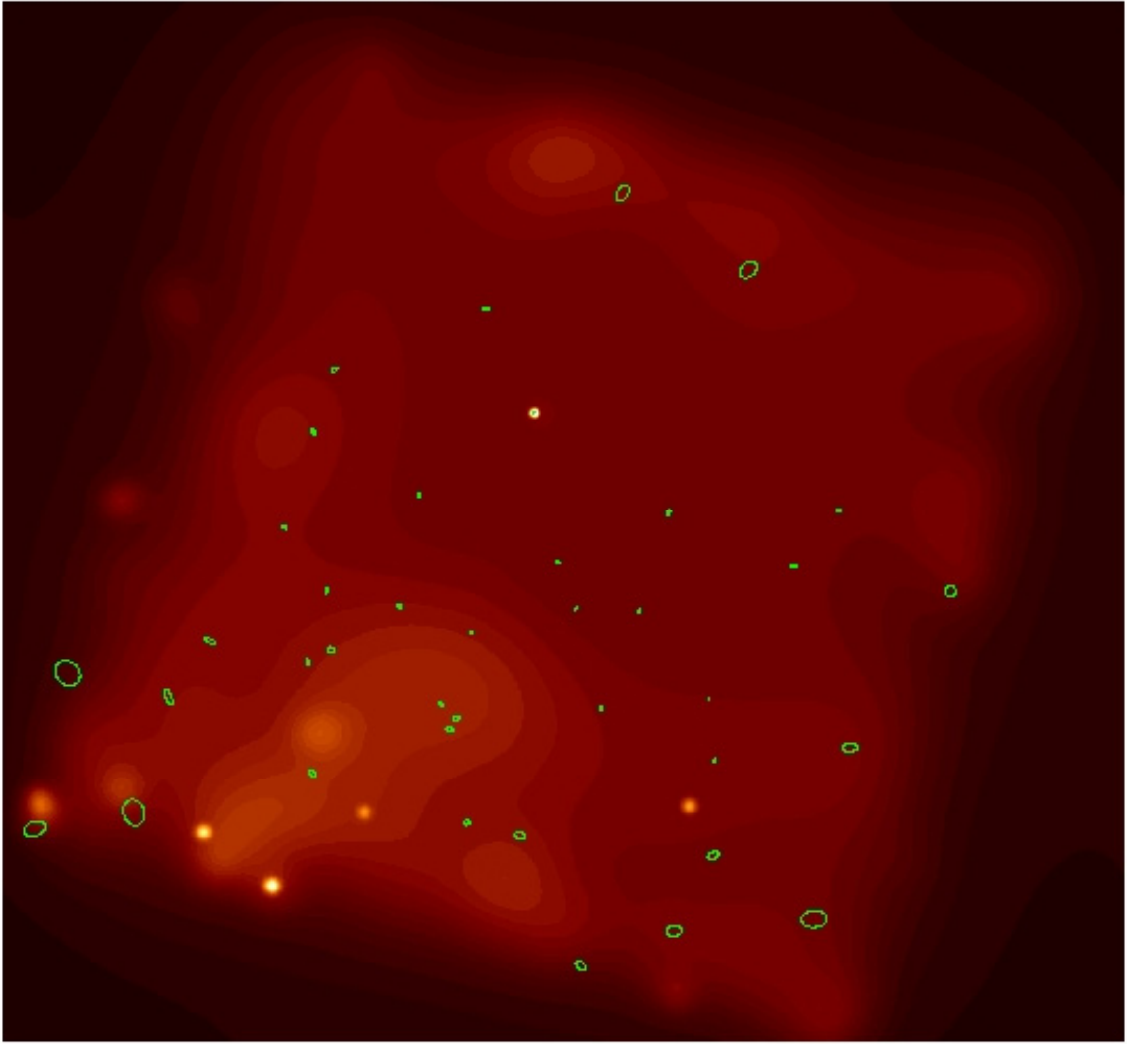

Fig. 3: An adaptively smoothed X-ray image of the TeV source region, covering the same field as in Fig. 2. The point-like sources have been removed prior to the smoothing – they are overlaid as the faint green contours. Some spurious maxima in the diffuse emission are artifacts of the smoothing algorithim. The suprious maxima are those which appear point-like, but have no true point-like (green contours) counterparts. eg. the two point-like maxima in the SE. North is up and East is to the left.

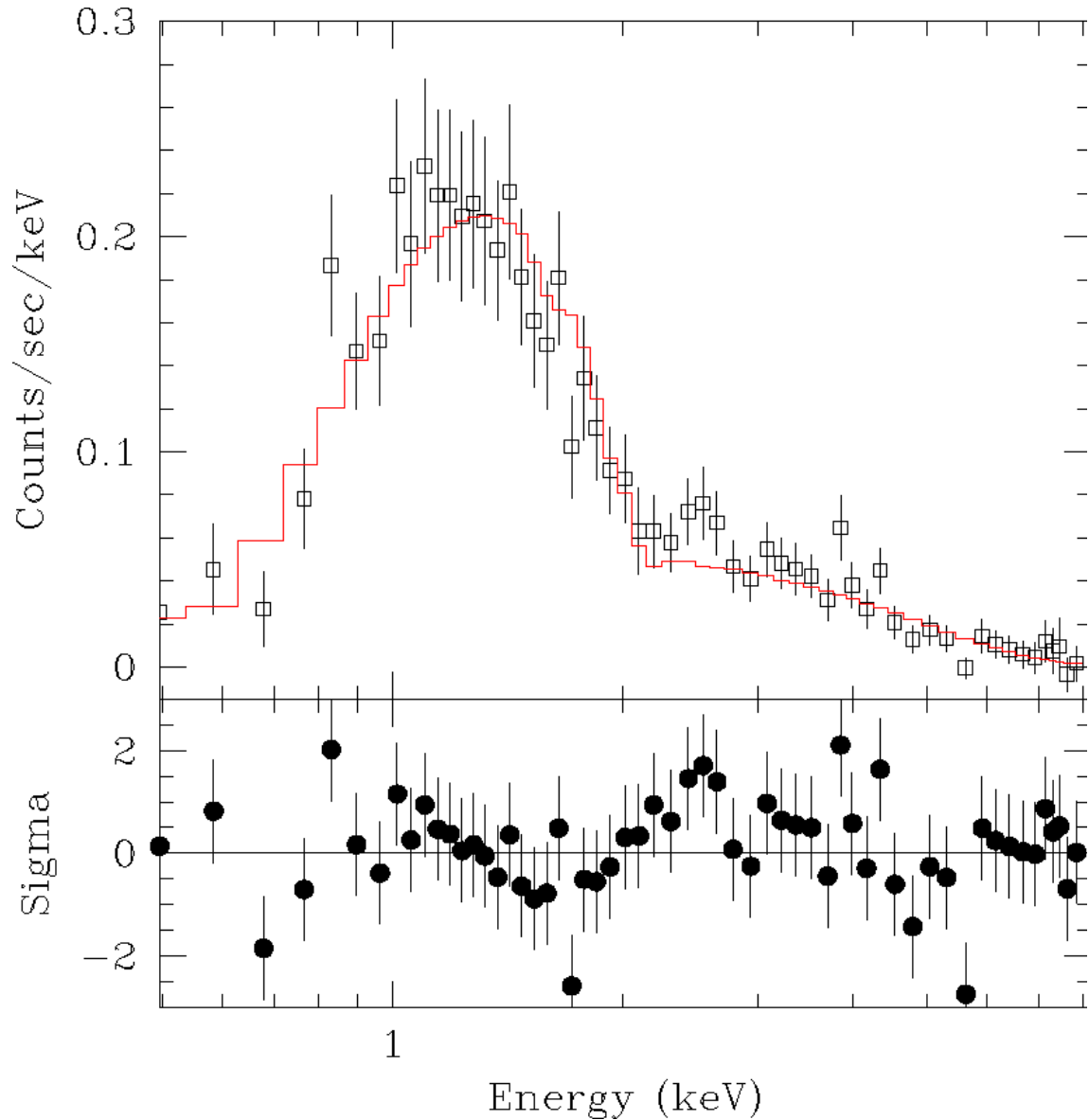

Fig. 4: ACIS pulse-height spectrum of the diffuse emission in the TeV source region and best-fit optically-thin plasma model, together with residuals in terms of $\sigma$. While there appear to be some systematic residuals, between 1 and 2 keV for example, the data are in general well-represented by the model, yielding a reduced $\chi^2$ of 0.9. However, due to the poor statistics we cannot discriminate between a thermal *vs.* non-thermal model in the short, 5ksec, integration. *The power-law fit also yielded a reduced $\chi^2$ of 0.9. Since the fit is qualitatively identical it is not shown here.*

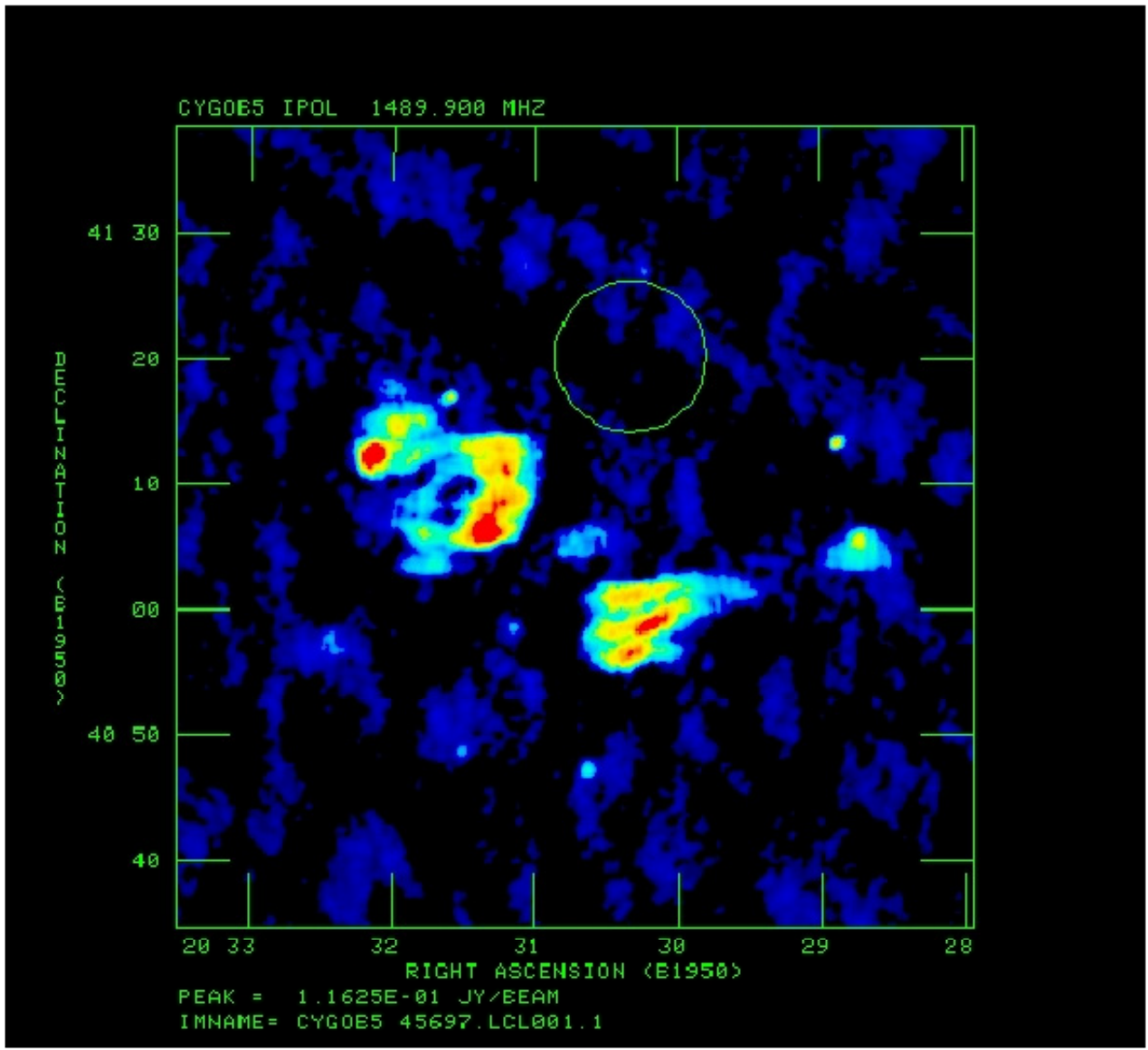

Fig. 5: The VLA D-configuration radio image of the Cyg OB2 region. The green circle shows the 5.6′ radius extent of the diffuse TeV source TeV J2032+4131 reported by HEGRA (Rowell et al. 2002; Aharonian et al., 2002). The upper limit to the radio emission there at 1.49 GHz is <200 mJy.

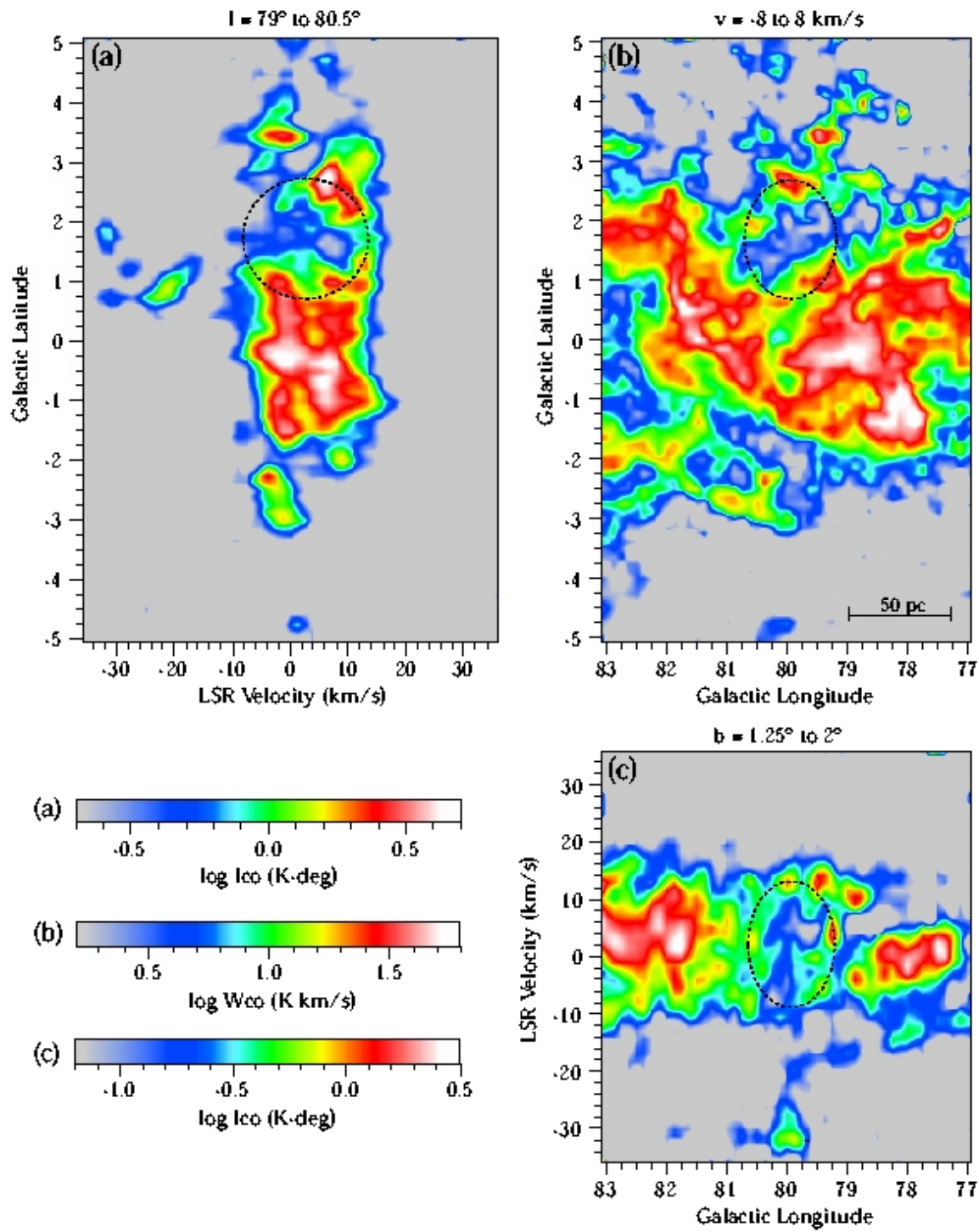

Fig. 6: The CO($J$=1→0) emission maps showing three orthogonal cuts through the $l$-$b$-$v_{lsr}$ data-cube. There is good evidence for an expanding cavity centered approximately $l,b$~(80.5,+1.8) in the velocity interval $v_{lsr}$ ~ -8 to +13 km sec$^{-1}$. The dotted ellipse is simply a by-eye fit to the 3 dimension of the shell. The $l$-$b$ map also shows evidence for other partial shells roughly centered on Cyg OB2 (mainly toward lower latitude). In addition, the $b$-$v_{lsr}$ and $l$-$v_{lsr}$ maps suggest that a front section of the shell may have been blown out toward us, the remnants of that section perhaps seen at $v_{lsr}$ ~ −30 km/s.

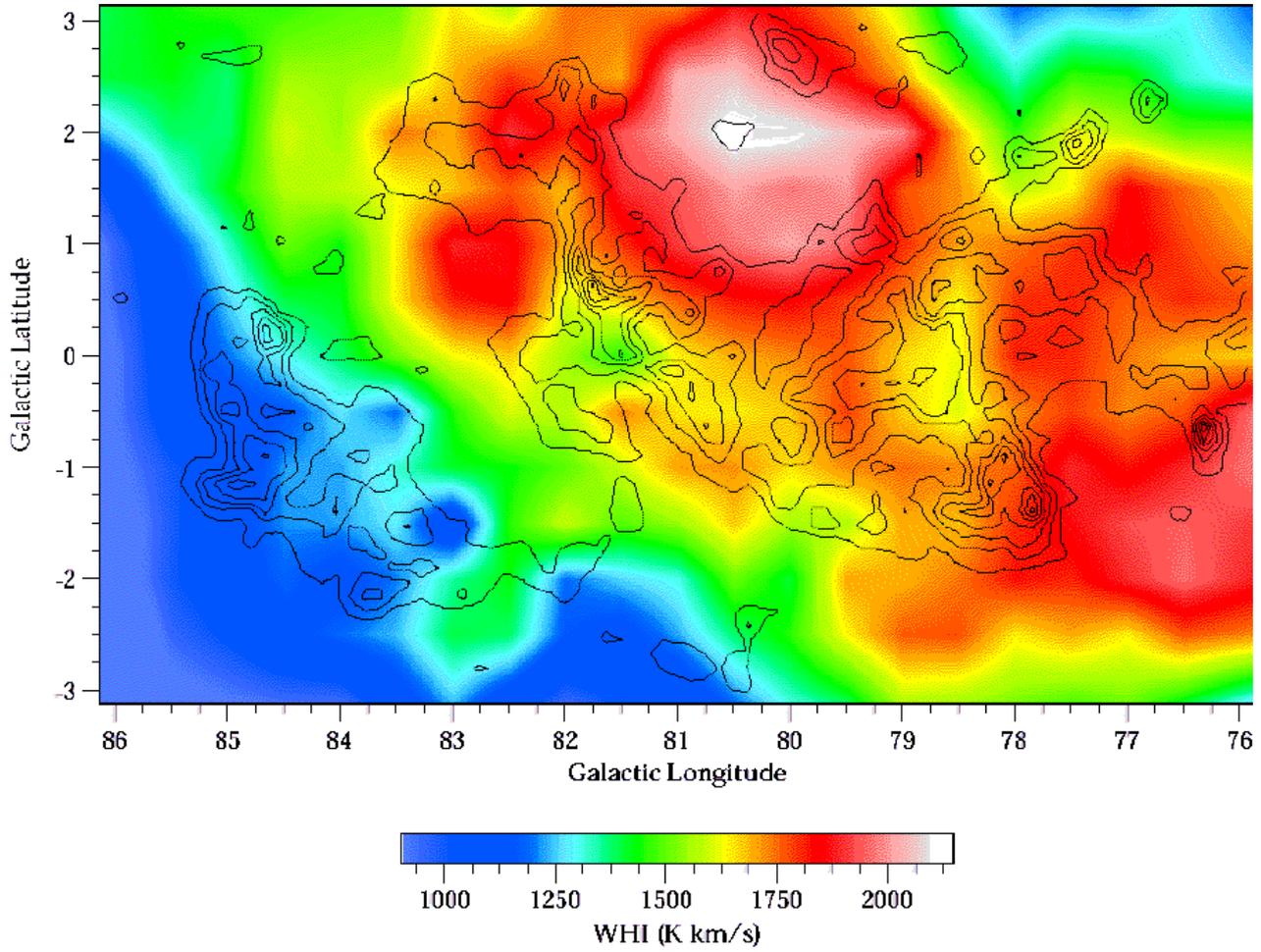

Fig. 7: Similar to Fig 6, but here the color scale is HI intensity (21 cm emission integrated -6 to 10 km sec$^{-1}$), and the contours are CO integrated over the same range, tracing the H$_2$ column density. Since the CO partial shell (centered $l,b$~80.5,+1.8) encloses the HI (in $l$-$v$ and $b$-$v$ space also; see Fig 8a&b), a reasonable interpretation is that the ambient molecular hydrogen is being disassociated by the expanding shell. Note that Langston et al. (2000) have found a number of HII regions located at the periphery of the shell-like structure (see text).

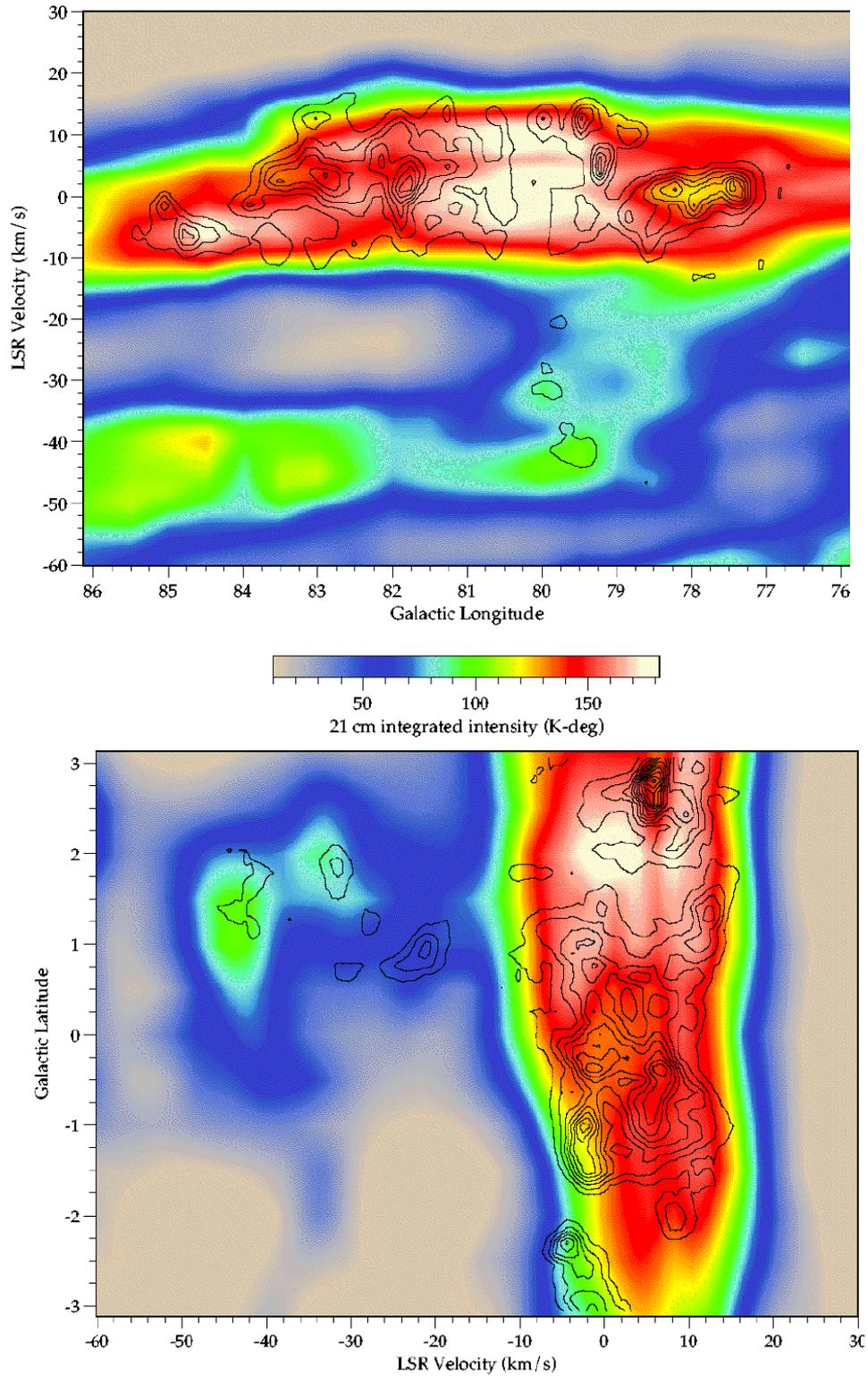

Fig. 8: Similar to Fig 7, but here the two panels show the two other orthogonal cuts through the HI (color) and CO (contours) data-cubes: (a) *l-v* map integrated over the range *b*=1° to 2°; CO contour spacing is 0.5K-deg, starting at 0.5 K-deg (b) the *b-v* map integrated over the range *l*=79.5° to 80.5°; CO contour spacing is 0.4 K-deg, starting at 0.4 K-deg. Note how the CO shell seen near *l*~80° in (a) and near *b*~1.8° in (b) coincides in velocity with an HI enhancement.

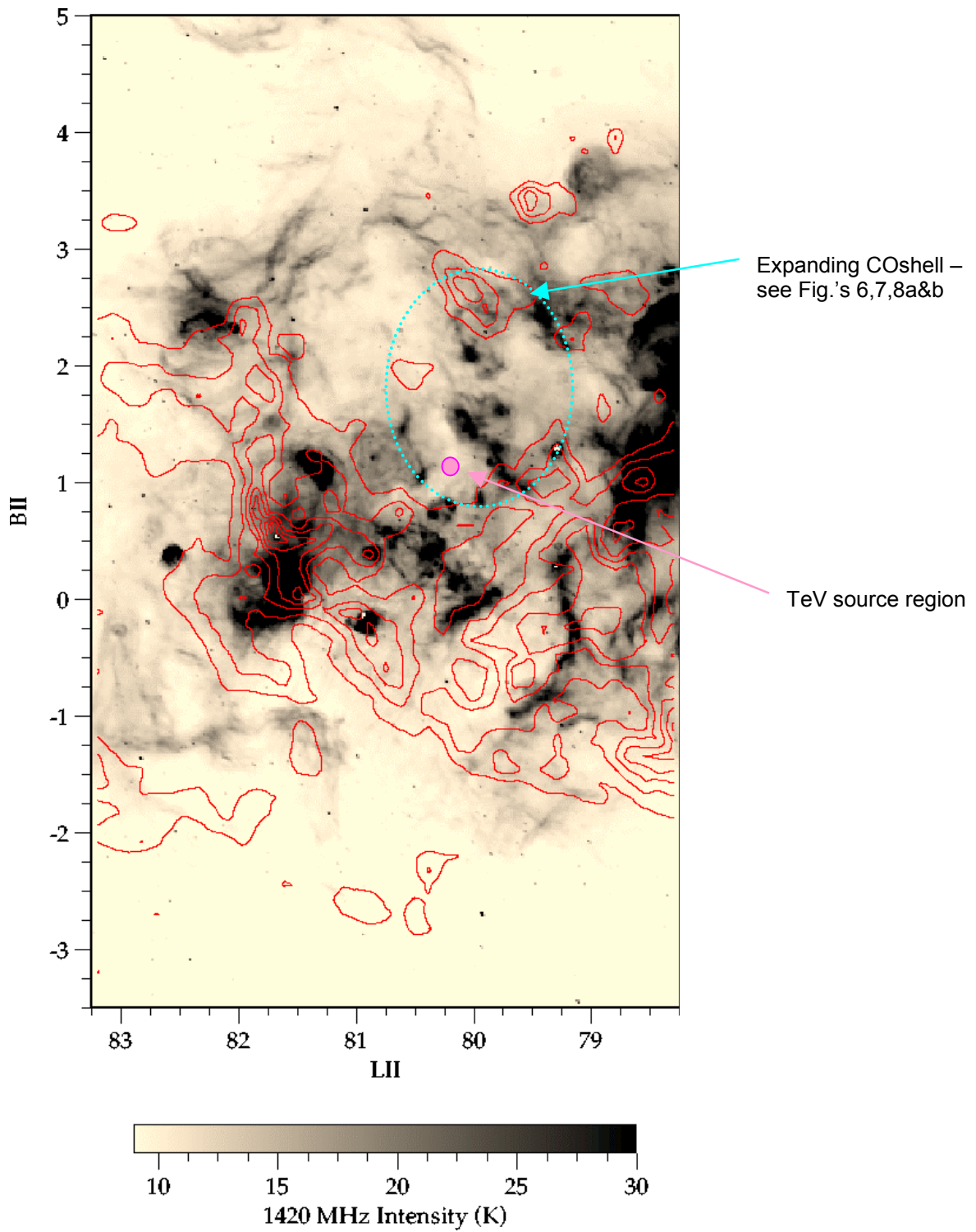

Fig. 9: The CO contours (-6 to 10 km sec$^{-1}$) are shown overlaid on a 1.420 GHz intensity map obtained from the Canadian Galactic Plane Survey. The locations of the expanding shell (see Fig.'s 6,7,8a&b) and the TeV source are marked. Note the possible relationship between the CO distribution and the radio structures in the region near $l,b \sim 80.5,+1.8$.

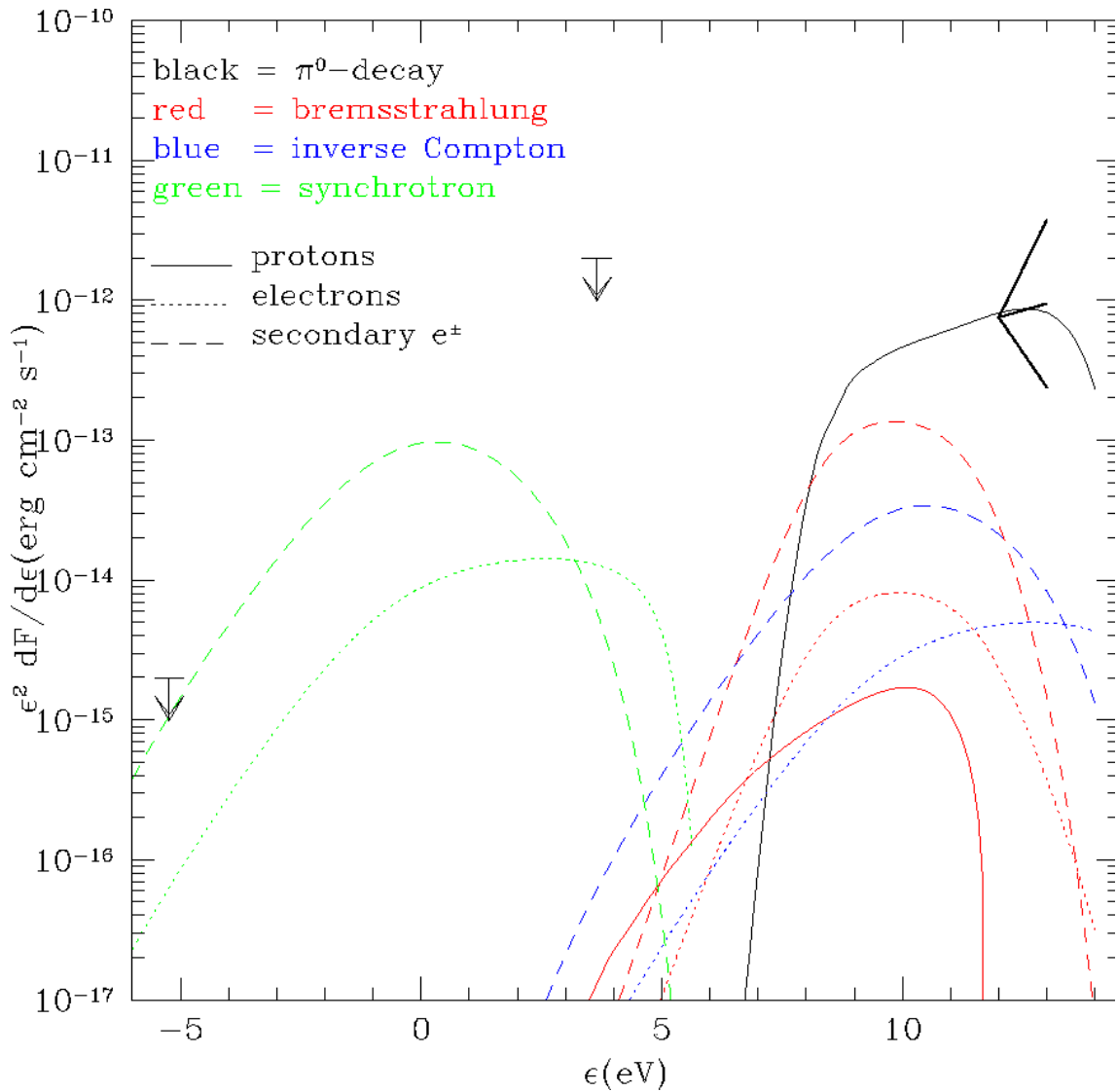

Fig. 10: A simulated multiwavelength spectrum for the case where the source TeV J2032+4131 has a predominantly hadronic origin. The ratio of primary electrons to protons was taken as 1%. A weak magnetic field of 5μG was assumed, in line with the nominal Galactic value. Interestingly, the radio emission of the secondary electrons dominates the contribution from the primaries – this is because the age of the source (~2.5Myrs) exceeds the cooling time of the secondary $e^\pm$ and thus they simply accumulate in the source region. The injection efficiency (ratio of GCR energy to time-integrated wind power) is 0.08%. *Note that the measured X-ray flux is taken here as an upper limit to the non-thermal component alone. Deeper X-ray and radio observations will help resolve the diffuse non-thermal components, which could then be directly compared with the simulated spectrum shown here.*

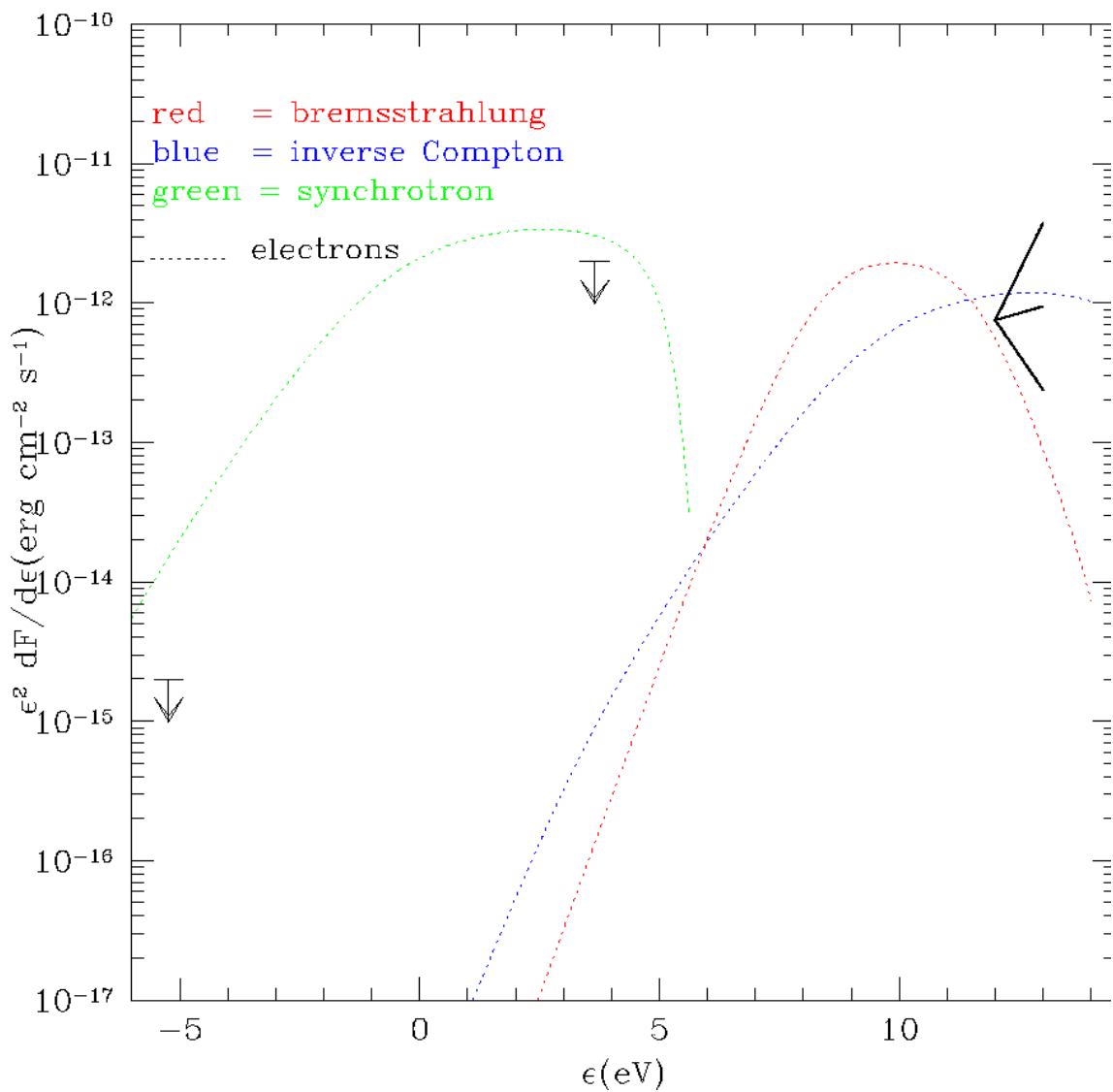

Fig. 11: A simulated multiwavelength spectrum for the case where the source TeV J2032+4131 has a purely electronic origin. A weak magnetic field of 5$\mu$G was assumed, in line with the nominal Galactic value. The injection efficiency (ratio of required GCR energy to time-integrated wind power) in this case is 0.2%. *Note that since both the X-ray and radio upper limits are violated and thus an electronic origin of TeV J2032+4131 is disfavored.* If a lower magnetic field exists in the TeV source region this would, of course, decrease the synchrotron emission (green curve), and could allow for an electronic model. However, Crutcher & Lai (2002) find that magnetic fields in young star forming regions are typically even higher – and not lower – than the nominal Galactic value of 5$\mu$G we have used here. *Note that, as in the previous figure, the measured X-ray flux is taken here as an upper limit to the non-thermal component alone. Deeper X-ray and radio observations will help resolve the diffuse non-thermal components, which could then be directly compared with the simulated spectrum shown here.*

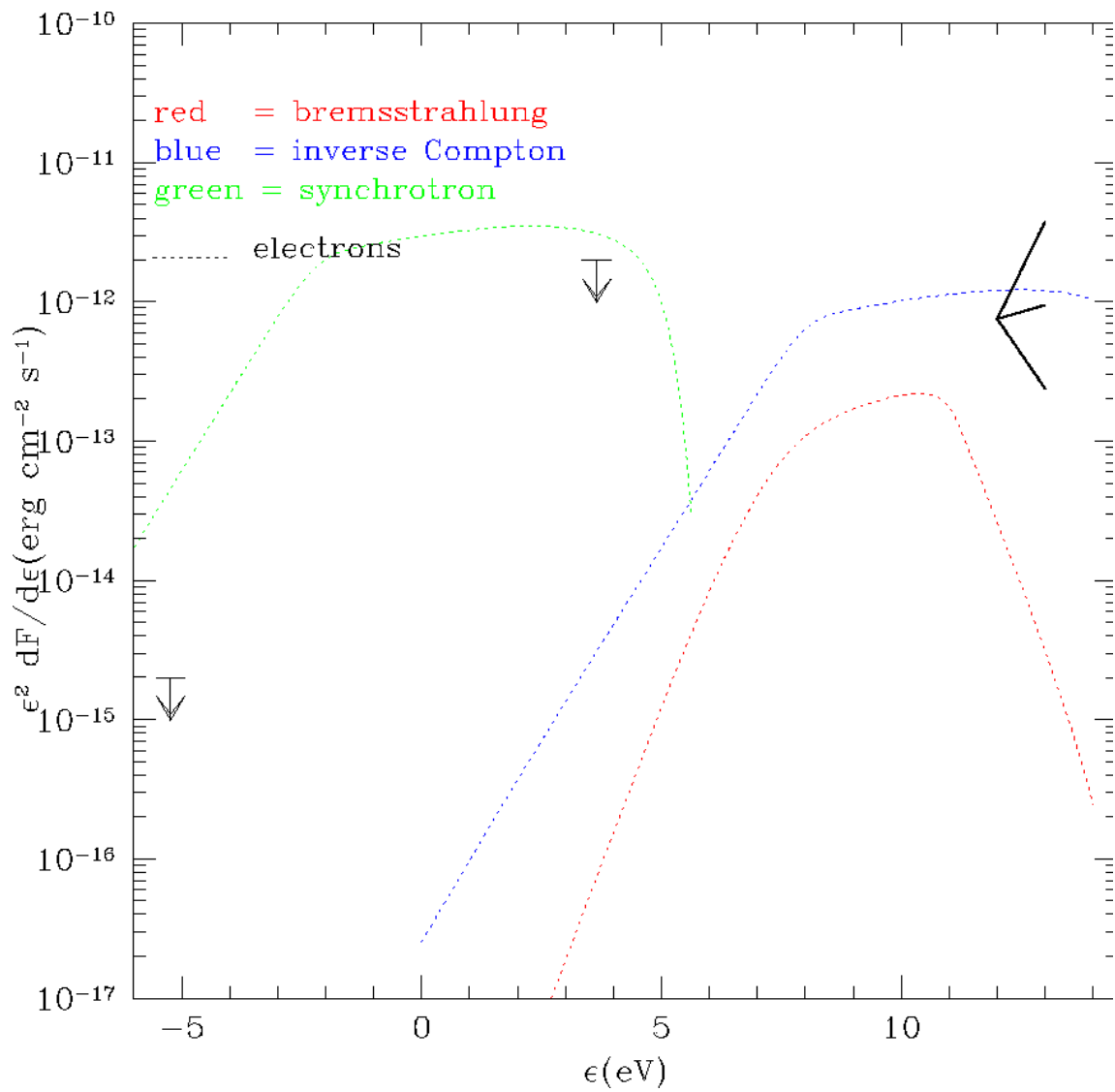

Fig. 12: Same as Fig. 11, except a density of 1cm$^{-3}$ is used instead of the empirically determined ~30cm$^{-3}$.